\documentclass[paper,11pt]{JHEP}
\usepackage{amsfonts}
\usepackage{amsmath}
\usepackage{amssymb}
\usepackage{extarrows}
\usepackage{multirow}
\usepackage{cite}
\usepackage{inputenc}
\usepackage{xspace}
\usepackage{url}
\usepackage{mathtools}

\newcommand{\bra}{\langle}
\newcommand{\ket}{\rangle}

\newcommand{\be}{\beta}

\newcommand{\D}{\Delta}

\newcommand{\m}{\mu}
\newcommand{\n}{\nu}

\newcommand{\OO}{{\cal O}}

\newcommand{\beq}{\begin{equation}}
\newcommand{\eeq}{\end{equation}}

\newcommand{\mycomment}[1]{}

 \def\be{\begin{equation}}
\def\ee{\end{equation}}
\def\bea{\begin{eqnarray}}
\def\eea{\end{eqnarray}}

 \def\IZ{\relax\ifmmode\mathchoice
 {\hbox{\cmss Z\kern-.4em Z}}{\hbox{\cmss Z\kern-.4em Z}}
 {\lower.9pt\hbox{\cmsss Z\kern-.4em Z}}
 {\lower1.2pt\hbox{\cmsss Z\kern-.4em Z}}\else{\cmss Z\kern-.4em Z}\fi}
 \def\IB{\relax{\rm I\kern-.18em B}}
 \def\IC{{\relax\hbox{$\inbar\kern-.3em{\rm C}$}}}
 \def\Ic{{\relax\hbox{$\inbar\kern-.22em{\rm c}$}}}
 \def\ID{\relax{\rm I\kern-.18em D}}
 \def\IE{\relax{\rm I\kern-.18em E}}
 \def\IF{\relax{\rm I\kern-.18em F}}
 \def\IG{\relax\hbox{$\inbar\kern-.3em{\rm G}$}}
 \def\IGa{\relax\hbox{${\rm I}\kern-.18em\Gamma$}}
 \def\IH{\relax{\rm I\kern-.18em H}}
 \def\II{\relax{\rm I\kern-.18em I}}
 \def\IK{\relax{\rm I\kern-.18em K}}
 \def\IP{\relax{\rm I\kern-.18em P}}

\def\Tr{{\rm Tr}}
 \font\cmss=cmss10 \font\cmsss=cmss10 at 7pt
 \def\IR{\relax{\rm I\kern-.18em R}}

\def\bra{\langle}
\def\ket{\rangle}

\def\D{\Delta}

\def\m{\mu}
\def\n{\nu}

\renewcommand{\@}[1]{\sqrt{#1}}
\renewcommand{\le}[1]{\label{#1}\end{eqnarray}}

\def\ffract#1#2{\raise .35 em\hbox{$\scriptstyle#1$}\kern-.25em/
\kern-.2em\lower .22 em \hbox{$\scriptstyle#2$}}

\newdimen\tableauside\tableauside=1.0ex
\newdimen\tableaurule\tableaurule=0.4pt
\newdimen\tableaustep
\def\phantomhrule#1{\hbox{\vbox to0pt{\hrule height\tableaurule width#1\vss}}}
\def\phantomvrule#1{\vbox{\hbox to0pt{\vrule width\tableaurule height#1\hss}}}
\def\sqr{\vbox{%
  \phantomhrule\tableaustep
  \hbox{\phantomvrule\tableaustep\kern\tableaustep\phantomvrule\tableaustep}%
  \hbox{\vbox{\phantomhrule\tableauside}\kern-\tableaurule}}}
\def\squares#1{\hbox{\count0=#1\noindent\loop\sqr
  \advance\count0 by-1 \ifnum\count0>0\repeat}}
\def\tableau#1{\vcenter{\offinterlineskip
  \tableaustep=\tableauside\advance\tableaustep by-\tableaurule
  \kern\normallineskip\hbox
    {\kern\normallineskip\vbox
      {\gettableau#1 0 }%
     \kern\normallineskip\kern\tableaurule}%
  \kern\normallineskip\kern\tableaurule}}
\def\gettableau#1 {\ifnum#1=0\let\next=\null\else
  \squares{#1}\let\next=\gettableau\fi\next}

\newcommand{\torino}{Dipartimento di Fisica, Università di Torino
and Istituto Nazionale di Fisica Nucleare - sezione di Torino
Via P. Giuria 1 I-10125 Torino, Italy.}
\newcommand{\bangkok}{Department of Physics, Faculty of Science, Chulalongkorn University,
Thanon Phayathai, Pathumwan, Bangkok 10330, Thailand.}
\newcommand{\auth}{Institute of Theoretical Physics, Aristotle University of Thessaloniki, 54124 Thessaloniki, Greece.}

\newcommand{\pasadena}{Walter Burke Institute for Theoretical Physics,
 California Institute of Technology, Pasadena, CA  91125}
\newcommand{\UCLA}{Mani L. Bhaumik Institute for Theoretical Physics, 
Department of Physics and Astronomy, UCLA, Los Angeles, CA 90095}
\newcommand{\torvergata}{I.N.F.N. Sezione di Roma Tor Vergata, Via della 
Ricerca Scientifica 00133 Roma, Italy}
\allowdisplaybreaks
\title{The analytic structure of conformal blocks  \\ and the generalized Wilson-Fisher fixed points}
\author{Ferdinando Gliozzi $^a$, Andrea L. Guerrieri $^{b,c}$,
\,Anastasios C. Petkou $^d$ and \,Congkao Wen $^{e,f}$\\
$^a$\torino\\
$^b$\bangkok\\
$^c$\torvergata \\
$^d$\auth\\
$^e$\pasadena\\
$^f$\UCLA}

\abstract{
We describe in detail the method used in our previous work arXiv:1611.10344 to study the Wilson-Fisher critical points nearby generalized free CFTs, exploiting the analytic structure of conformal blocks as functions of the conformal dimension of the exchanged operator. Our method is equivalent to the mechanism of conformal multiplet 
recombination set up by null states. We compute, to the first non-trivial order in the $\epsilon$-expansion, the anomalous dimensions and the OPE coefficients of infinite classes of scalar local operators using just CFT data. We study single-scalar and $O(N)$-invariant theories, as well as theories with multiple deformations. When available we agree with older results,  but we also produce a wealth of new ones. Unitarity and crossing symmetry are not used in our approach and we are able to apply our method to non-unitary theories as well. Some implications of our results for the study of the non-unitary theories containing partially conserved higher-spin currents are briefly mentioned.
 }
\preprint{CALT-TH-2017-009}

\begin{document}
\maketitle

\section{Introduction}
The notion of criticality and its intimate relationship with phase transitions is central in our quests for understanding the physical world. Over the  past few decades,  significant progress in the study of criticality has been achieved for systems that can be described by quantum fields. In this case, critical behaviour is generally associated with the existence of conformal field theories (CFTs). The latter theories posses a large spacetime symmetry that allows the calculation of various physically relevant quantities such as scaling dimensions, coupling constants and central charges. This program has led to some remarkable results for two-dimensional systems where one can explore the infinite dimensional
Virasoro algebra \cite{Belavin:1984vu}.

Nevertheless, critical systems are also abundant in dimensions $d>2$ and therefore we are forced to study higher-dimensional CFTs to understand them better. Higher-dimensional CFTs are much harder to explore than their two-dimensional counterparts, and that explains the relatively slow progress in their study up until the end of the millennium. This has changed dramatically by the advent of AdS/CFT \cite{Maldacena:1997re,Gubser:1998bc,Witten:1998qj} that put the focus back into higher-dimensional CFTs and their relevance not only for critical systems, but for quantum gravity and string theory as well. It is inside this fertile environment of general rethinking about CFTs in $d>2$ that the more recent significant progress of the higher-dimensional conformal bootstrap \cite{Ferrara:1973yt} was born. The latter can be described as a combination of analytic and numerical tools that give remarkably accurate results for  critical points in diverse dimensions 
\cite{Rattazzi:2008pe,Rychkov:2009ij, Rattazzi:2010gj,Poland:2010wg,ElShowk:2012ht,Liendo:2012hy,
Pappadopulo:2012jk,ElShowk:2012hu,Gliozzi:2013ysa, Gaiotto:2013nva, El-Showk:2014dwa,
Beem:2013qxa,Nakayama:2014yia,Nakayama:2014sba,Gliozzi:2014jsa,Chester:2014fya, 
Kos:2014bka,Chester:2014gqa,Beem:2014zpa,Simmons-Duffin:2015qma, 
Bobev:2015vsa,Kos:2015mba,Bobev:2015jxa,Gliozzi:2015qsa,
Beem:2015aoa,Nakayama:2016jhq,Kos:2016ysd,Nakayama:2016cim,Gliozzi:2016cmg,Esterlis:2016psv,El-Showk:2016mxr}.

One of the intriguing features of the conformal bootstrap approach is that some properties of the critical systems that are strictly related to 
the renormalization group  description of phase transitions, can  be seen instead as a direct 
consequence of conformal invariance. For instance, in the quantum field theory 
 approach to the 
Wilson-Fisher (WF) \cite{Wilson:1971dc} fixed point of the $\phi^4$ theory in $d=4-\epsilon$ (or its generalizations in other dimensions with different 
marginal perturbations\footnote{Strictly speaking they are relevant deformations for $\epsilon>0$, we will often loosely call them marginal in the sense of $\epsilon \sim 0$.}) there are  two distinct small dimensionless parameters in the game: 
the coupling constant $g$ which turns on the interaction in the Lagrangian, and $\epsilon=4-d$. One performs a (scheme-dependent) loop expansion in 
$g$ of some physical quantity like for instance the anomalous dimensions of local operators. When $d$ is slightly smaller than four the perturbation $\phi^4$ becomes slightly relevant at the Gaussian UV fixed point  and the system flows to the infrared WF fixed point. The vanishing of the Callan-Zymanzik $\beta(g)$ equation fixes the relation between $g$ and $\epsilon$ and gives 
scheme-independent $\epsilon$-expansions for the anomalous dimensions. 

The conformal bootstrap approach tells a parallel but different story. 
For instance, in the numerical bootstrap it suffices to ask for a 
consistent CFT with a scalar field 
$\phi$ having an operator product expansion (OPE) $[\phi]\times[\phi]=1+[\phi]+\dots$, in order to be able to select a unique solution of the bootstrap equations in the whole range $2<d<6$. This procedure allows one to evaluate the low-lying spectrum of 
the primary operators of a $\phi^3$ theory \cite{Gliozzi:2014jsa} which compares well with strong coupling expansions and Monte Carlo 
simulations as well as  with more recent conventional $\epsilon$-expansions \cite{Gracey:2015tta}. If one restricts the search to {\it unitary} CFTs where convex optimization methods apply, one  is led to a wealth of non-trivial results which 
are particularly impressive in the case of 3d Ising model, where very precise determinations of the bulk critical exponents are obtained \cite{El-Showk:2014dwa,Kos:2014bka,Kos:2016ysd}.  However, the analytic approaches to bootstrap, 
if we exclude supersymmetric theories, have not yet reached the high level
of accuracy of the conventional  $\epsilon$-expansion in quantum field theory,  but nevertheless are in many cases much more simple to apply and sometimes give results that cannot be obtained by other analytic methods. For instance in the recent approach where the 4pt functions are expanded in terms of exchange Witten diagrams instead of the 
conventional conformal blocks \cite{Gopakumar:2016wkt,Gopakumar:2016cpb,Dey:2016mcs}, one obtains anomalous dimensions of some local operators  to $O(\epsilon^3)$.  Similarly in a study of CFTs with 
weakly broken higher-spin symmetry the spectrum of broken currents is obtained at the first non-trivial order of the breaking parameter \cite{Skvortsov:2015pea,Giombi:2016hkj,Alday:2016njk,Alday:2016jfr,Bashmakov:2016uqk,Manashov:2016uam,Giombi:2016zwa,Giombi:2017rhm}. The Wilson-Fisher points with $O(N)$ invariant symmetry have been studied in the limit of large global charges in \cite{Hellerman:2015nra, Alvarez-Gaume:2016vff}. In \cite{Rychkov:2015naa}, using the fact that in a $\phi^4$ 
theory in $4-\epsilon$ dimensions  the equations of motion imply that 
$\phi^3$ becomes a descendant of $\phi$ at the WF fixed point, the anomalous dimensions of operators are 
obtained with no input from perturbation theory. Such an approach has also been generalized in various ways  in \cite{Basu:2015gpa, Nii:2016lpa,Hasegawa:2016piv,Bashmakov:2016uqk,Roumpedakis:2016qcg, Liendo:2017wsn}. 

In a recent short paper  \cite{Gliozzi:2016ysv} we have extended some of the above ideas to the wide class of  generalized free CFTs. 
More specifically our method is based on considering two of the three axioms of
\cite{Rychkov:2015naa}, but we did not have to assume any Lagrangian equations of motion (eom). 
We pointed out that the mechanism of recombination of conformal multiplets\footnote{Conformal multiplet recombination has been discussed in the context of the holography of higher-spin theories in \cite{Girardello:2002pp,Leigh:2003gk}. For a more recent work see \cite{Bashmakov:2016pcg}.}
can be directly read from the 
analytic properties of the conformal blocks without further assumptions. 
We considered in particular the set of scalar states with dimensions 
$\Delta_k=\frac d2-k$ where $k=1,2,\dots$; $k=1$ corresponds to the dimension 
of a canonical scalar field, while  $k>1$ corresponds to the 
subclass of generalized free CFTs with $\Box^{k}$ kinetic term that are coupled to the stress 
tensor \cite{Osborn:2016bev,Brust:2016gjy}. These states have generically scalar null descendants with dimensions
$\Delta_{\rm desc}=\frac d2+k$, which is equivalent to saying that their conformal blocks are in principle singular at $\Delta=\Delta_k$. However, since the  4pt functions of the free theory are
always finite, the states with $\Delta_k=\frac{d}{2}-k$ must {\it either} decouple in the free field theory limit {\it or} have their singularities somehow removed. In the first case it is the descendants that emerge as  regular primary fields in the free theory limit. In the second case, as it was firstly discussed in \cite{Guerrieri:2016whh}, the would-be singular blocks with $\Delta_k=d/2-k$ do arise in the OPE of  {\it free} nonunitary CFTs but their  singularities cancel out due to the presence of corresponding null states in the spectrum. The fate of singular  blocks in the conformal OPE is briefly summarized in Appendix B.  

 In this work we will heavily use the first of the above two mechanism in order to calculate explicitly various critical quantities in nontrivial CFTs. As we switch-on the interaction 
by assuming non-vanishing anomalous dimensions in $d-\epsilon$, we have $\Delta_k\neq \frac{d}{2}-k$ and nothing is singular anymore. In a CFT language the corresponding scalar multiplet becomes long. Using then the explicit results for the residues of the conformal multiplets, and the known OPE coefficients of the free theory, we will be able to calculate analytically the leading corrections to the anomalous dimensions for wide classes of operators in many nontrivial CFTs.  In particular,  we find that for any 
pair of positive integers $k,n$ the (generically fractional) space dimension $d=2nk/(n-1)$ is an  
upper critical dimension, 
in the sense that there is a consistent smooth deformation 
of the free theory at $d-\epsilon$ representing a {\it generalized}  WF fixed 
point associated with the marginal perturbation $\phi^{2n}$. At such  WF fixed points we can calculate the anomalous dimensions of composite operator of the form $\phi^p$ with $p\in \mathbb{Z}^+$, as well as non-trivial OPE coefficients. We are also able to present $O(N)$-symmetric generalizations of the above WF fixed points, and calculate the anomalous dimensions, at the first non-trivial order in $\epsilon$, of the corresponding scalar composite operators $\phi^p$ as well as of fields carrying symmetric traceless tensors representations of $O(N)$, having general form $\phi^{(s)}_{i_1\dots i_s}\phi^{2p}$, for any $p$ and spin $s$. 
In this work we describe in more detail our calculations presented in \cite{Gliozzi:2016ysv} and present some new results.

The content of this paper is as follows.
In Section \ref{sec:singularities} we study with a new method the singularities of generic conformal blocks as a function of the scaling dimension $\D$ of the exchanged operators in a 
4pt function of four arbitrary scalars. Requiring the cancellation  of singularities in the expansion in conformal blocks of a suitable function we find that 
conformal blocks have simple poles in $\D$.
The position and corresponding residues of  
these poles  coincide with those dictated by the null states and obtained in  
\cite{Kos:2014bka, Penedones:2015aga} by completely different methods. Here we also point out the intimate relationship of the poles in generic conformal blocks with  partially conserved higher-spin currents and their corresponding generalized Killing tensors. 
In Section \ref{sec:GFCFT} we review and improve the analysis of \cite{Guerrieri:2016whh} 
regarding the definition and the OPE of  generalized free CFTs. Also, generalizing our earlier results, we give a remarkably simple formula for the total central charge of nonunitary generalized free CFTs in arbitrary even dimensions.
 In Section \ref{sec:deformations} we  study in detail the {\it generalized} Wilson-Fisher  fixed points near  
generalized free CFTs in arbitrary dimensions for single-scalar theories and give analytic results for the anomalous dimensions of large classes of operators and OPE coefficients to leading order in the $\epsilon$-expansion.  In sections \ref{subsection:O(N)} and section \ref{subsection:d=6} we study respectively  nontrivial CFTs  with O$(N)$ global symmetry and theories with multiple marginal deformations. We give there too results for anomalous dimensions and OPE coefficients that match earlier calculations, but we also give many new ones. Finally in Section \ref{sec:conclusion}  we draw some conclusions. Technical details of our calculations are presented in the Appendices. 
  
\section{The analytic structure of the conformal OPE}
\label{sec:singularities}
\subsection{Singularities of generic conformal blocks}
We begin by studying the analytical structure of generic conformal blocks as functions of the conformal dimension $\Delta$ of the exchanged operator. This will play the central role in our study of WF fixed points in this paper. For our purposes we will consider conformal blocks with general external scalar operators. In a generic  CFT in $d$ dimensions the 4pt function of arbitrary scalars
can be parametrised as \cite{Hogervorst:2013kva}
\be
\bra \OO_1(x_1)\OO_2(x_2)\OO_3(x_3)\OO_4(x_4)\ket=\frac{g(u,v)}
{\vert x_{12}\vert^{\Delta^{+}_{12}}
\vert x_{34}\vert^{\Delta^{+}_{34}}} \left( {\vert x_{24}\vert \over \vert x_
{14}\vert} \right)^{\Delta^{-}_{12}}
\left( {\vert x_{14}\vert \over \vert x_{13}\vert} \right)^{\Delta^{-}_{34}} 
\,,
\label{4pt}
\ee
where $\Delta^{\pm}_{ij} = \Delta_{i} \pm \Delta_{j}$, $i,j=1,2,3,4$ and $\Delta_{i}$ is the
 scaling dimension of $\OO_i$, while $u={x_{12}^2 x_{34}^2 \over x_{13}^2 x_{
24}^2}$ and $v={x_{14}^2 x_{23}^2 \over x_{13}^2 x_{24}^2}$ are the cross ratios. 
The function $g(u,v)$ can be expanded in terms of conformal blocks
$G^{a,b}_{\Delta,\ell}(u,v)$, i.e. eigenfunctions of the quadratic (and quartic) 
Casimir operators of $SO(d+1,1)$:
\be
g(u,v)=\sum_{\Delta,\ell}p_{\Delta,\ell} \,G^{a,b}_{\Delta,\ell}(u,v)\,,\,\,\,\,a=-\frac{\Delta^-_{12}}2\,,\,b=\frac{\Delta^-_{34}}2\,,
\ee
with 
$\Delta$ and $\ell$ being the scaling dimensions and the spin of the 
primary operators contributing to the $\{12\},\{34\}$
channel. The scalar 2pt functions are normalized as
\be
\label{eq:2pt}
\bra\OO_i(x_1)\OO_j(x_2)\ket=\frac{\delta_{ij}}
{\vert x_{12}\vert^{\Delta_i+\Delta_j}}\,,
\ee
and we normalize here  the conformal blocks  such that putting $u=z^2$ and $v=(1-z)^2$ in the limit $z\to0$ we have
\be
 G^{a,b}_{\Delta,\ell}(u,v)=z^\Delta+{\rm higher~order~terms}.
\label{eq:cbnorm}
\ee

These  conformal blocks form a complete 
basis which can be used to expand 
a general function of $u$ and $v$. Our starting point is the expansion of
$u^\delta$ into conformal blocks $G^{a,b}_{\Delta,\ell}(u,v)$, 
with $\delta,a,b$ and $\nu=\frac d2-1$  arbitrary  parameters.
We will use such an expansion to extract the positions and the residues of the poles of  the conformal blocks in the $\D$ variable. The same expansion will be used to find the OPE of generalized free field theories, however the results of this section are valid for any CFT in arbitrary dimensions.

A systematic method of finding the conformal block expansion of $u^\delta$ is described in detail in the Appendix \ref{appendix:expansion} and here we briefly sketch its salient features. We begin with the following expansion, 
\beq \label{eq:uexpansion}
u^\delta=\sum_{\Delta,\ell} \lambda^{a,b}_{\Delta,\ell}G^{a,b}_{\Delta, \ell}(u,v) \, ,
\eeq 
and the goal is to find the coefficient $\lambda^{a,b}_{\Delta,\ell}$. The idea is to apply the following differential operator
\beq
\Omega(n)=\prod_{[\D,\ell]\in\Sigma,2\tau+\ell< n}\left(C_2-c_2(\D,\ell)\right) \,, 
\eeq
to both sides of (\ref{eq:uexpansion}). $C_2$ is the  quadratic Casimir operator of $SO(d+1,1)$ while $c_2(\D,\ell)$ is its eigenvalue. The definitions and explicit formulae of $C_2$ and $c_2(\D,\ell)$ can be found in Appendix \ref{appendix:expansion}. $\Omega(n)$ projects out all the conformal blocks corresponding to the eigenvalues appearing in the product. Applying recursively $\Omega(n)$, as we show in Appendix \ref{appendix:expansion}, we obtain the following identity
\beq
u^\delta=\sum_{\tau=0}^\infty\sum_{\ell=0}^\infty\lambda^{a,b}_{\delta,\tau,\ell}G^{a,b}_{2\delta+2\tau+\ell, \ell}(u,v)= \sum_{\tau=0}^\infty\sum_{\ell=0}^\infty 
\frac{(-1)^\ell(2\nu)_\ell}{\tau!\ell!(\nu)_\ell(\nu+\ell+1)_\tau}
c^{a,b}_{\delta,\tau,\ell}
G^{a,b}_{2\delta+2\tau+\ell, \ell}(u,v),
\label{eq:cbexpansion}
\eeq 
where $(x)_y=\frac{\Gamma(x+y)}{\Gamma(x)}$ is the Pochhammer symbol.
The coefficient $ c^{a,b}_{\delta,\tau,\ell}$ is  given by
\beq
\label{eq:cab}
c^{a,b}_{\delta,\tau,\ell}=\frac{\prod_{i=a,b}(i+\delta)_{\ell +\tau}
(i+\delta-\nu)_\tau}{
(\Delta-1)_\ell(\Delta-\nu-\tau-1)_\tau(\Delta-2\nu-\tau-\ell-1)_\tau},
\eeq
with $\Delta=2\delta+2\tau+\ell$.
Clearly it  has three families of simple poles at $\Delta=\Delta_k$, with 
\bea \label{eq:singularities}
\Delta_k&=&1-\ell+k~~~(k=1, 2, \ldots,  \ell)\, , \cr
\Delta_k&=&1+\nu+k~~~(k=1, 2, \ldots, \tau) \, , \cr
\Delta_k&=&1+\ell+2\nu+k~~~(k=1, 2, \ldots, \tau) \, .
\eea 
The left-hand side of (\ref{eq:cbexpansion}), $u^{\delta}$, is a 
regular function of $\Delta$, thus such singularities must cancel on the 
right-hand side. Since the $G^{a,b}_{\Delta,\ell}$'s are linearly independent, 
the only way for such a cancellation to happen is that another conformal block $G^{a,b}_{\Delta', \ell'}(u,v)$ must appear in the sum, having  
$\Delta'=2\delta+2\tau'+\ell'$, and most importantly it becomes singular with the opposite sign residue, namely the residue should be proportional to yet another conformal block $G^{a,b}_{\Delta,\ell}$. Thus we can write
\bea
G^{a,b}_{\Delta', \ell'}(u,v) \sim {R^{a,b}(k,\ell) \over \Delta' - \Delta_k'}  G^{a,
 b}_{\D_k, \ell} (u, v) \, ,
\label{eq:position&residue}
\eea
with  $\Delta_k'- \Delta_k= \Delta'- \Delta$. We note $\Delta$ and $\Delta'$ actually differ by an integer, namely, $\Delta'-\Delta=\ell'-\ell+2\tau'-2\tau$. 
  
 Now another key observation is to recall that the  conformal blocks 
$G^{a,b}_{\D,\ell}(u,v)$  
are eigenfunctions not only of the quadratic  Casimir $C_2$ but {\it also} of the quartic Casimir operator $C_4$ of $SO(d+1,d)$ with corresponding eigenvalue $c_4(\Delta,\ell)$. Therefore  $G_{\Delta_k,\ell}(u,v)$ and 
$\lim_{\Delta'\rightarrow \Delta'_k}G_{\Delta', \ell'}(u,v)$ share the same eigenvalues
$c_2$ and $c_4$. We look for the possible solutions of the system of algebraic  equations
\bea \label{eq:casimireqs}
c_2(\Delta, \ell)=c_2(\Delta', \ell')\,, \quad c_4(\Delta, \ell )=
c_4(\Delta', \ell').
\eea 
where the Casimir eigenvalues are respectively, 
\begin{align} \label{eq:casimirequations}
c_2(\Delta, \ell) &= \frac{1}{2}\Delta ( \Delta -d ) + \frac{1}{2}\ell(\ell + d -2)\, ,\\
c_4(\Delta, \ell) &= \Delta^2 ( \Delta -d )^2 + {1 \over 2} d (d-1) \Delta (\Delta -d)+ \ell^2 (\ell+d -2)^2 + {1 \over 2} (d-1)(d-4)\ell (\ell+d-2) \, . \nonumber
\end{align}
By solving these  equations we discover a precise link between the 
position of the poles $\D'_k$ of a generic conformal block 
$G^{a,b}_{\D',\ell'}$ and the scaling dimension and spin 
$[\D_k,\ell]$ of the conformal block contributing to the residue in (\ref{eq:position&residue}). We explicitly find three possible families of solutions   shown in 
table \ref{tab:polepositions}. In the first 
and the third rows of the table 
the condition that $\D'_k-\D_k={\rm integer}$ follows automatically from the solution of the equations. In the second row, namely when $\ell'=\ell$, 
the constraint from quartic Casimir is actually redundant, so the condition  
that $\Delta'-\D$ ={\rm integer} is necessary to find the solutions. 
It is amusing to  see that we re-obtain in such a purely algebraic way the spectrum of the representations listed in (\ref{eq:singularities}). In other words, the requirement of the cancellation of singularities of the coefficients 
of the expansion (\ref{eq:cbexpansion}) with the poles of the conformal blocks 
can be equivalently reformulated as the search of solutions of the algebraic system (\ref{eq:casimirequations}) combined, in some cases, with the condition that $\D'-\D$ be an integer.  Furthermore requiring the complete cancellation of the singularities in (\ref{eq:cbexpansion}) allows us to fix the factor $R(k,\ell)$ in the residue for each case. In the following we will study each case separately and compute the corresponding $R(k,\ell)$. 

In the first case, namely $[\Delta_k'=1-\ell'-k, \, \ell']$ and 
$[\Delta_k=1-\ell+k, \, \ell=\ell'+k]$, the OPE coefficient in 
(\ref{eq:cbexpansion}) becomes singular for $k = 1, 2, \ldots, \ell$, 
and its residue is given by
\bea
r^{(1)}(k, \ell, \tau) = \frac{(-1)^k (2\nu)_\ell}
{\tau!\ell!(\nu)_\ell(\nu+\ell+1)_\tau}
 \frac{(a+\delta)_{\ell+\tau}(b+\delta)_{\ell+\tau}(a+\delta-\nu)_\tau(b+\delta-\nu)_\tau}{ \Gamma(k) \Gamma(\ell+1-k) (k-\ell-\nu-\tau)_\tau(k-2\nu-\tau-2\ell)_\tau} \, ,
\eea
\begin{table}
\centering
\begin{tabular}{c|c|c|cl}
$\Delta_k'$ & $\Delta_k$ & $\ell$  \\
\cline{1-4}
$1-\ell'-k$ & $1-\ell +k$ & $\ell'+k$    & \quad $k=1,2,\dots$\\
${d \over 2}-k$  & ${d \over 2}+k$  & $\ell'$     & \quad $k=1,2,\dots$\\ 
$d+\ell'-k-1$ & $d+\ell+k-1$ & $\ell'-k$   & \quad $k=1,2,\dots, \ell' $
\end{tabular}
\caption{Solutions of the algebraic system (\ref{eq:casimirequations}). The first column lists the position of the poles of a generic conformal block $G^{a,b}_{\D',\ell'}$. The second and third columns list the scaling dimension $\D_k$ and 
the spin $\ell$ 
of the conformal block contributing to the residue in (\ref{eq:position&residue}). }
\label{tab:polepositions}
\end{table}
with $\delta={1-2\ell-2\tau+k \over 2}$. The cancellation of the singularity requires that
\bea
r^{(1)}(k, \ell+k, \tau) = - R^{a,b}_1(k, \ell) \times \lambda^{a, b}_{\tau, \ell} \, . 
\eea
It thus yields
\bea
R^{a,b}_1(k, \ell)=  - { r^{(1)}(k, \ell+k) \over \lambda^{a, b}_{\tau, \ell} }
= -\frac{ k (-1)^k}{(k!)^2}\frac{(\ell+ 2 \nu)_{k} \left( {1-k  \over 2} + a\right)_{k} \left( {1-k  \over 2} + b\right)_{k}  }{(\ell+ \nu)_{k}} \, .
\label{eq:residue1}
\eea
We then move to the second case, the OPE coefficient becomes singular when 
$\Delta_k={d \over 2}+k$ for $k = 1, 2, \ldots, \tau$, and the residue is 
given by
\bea
r^{(2)}(k, \ell, \tau) = \frac{(-1)^{\ell + \tau + k}(2\nu)_\ell}
{\tau!\ell!(\nu)_\ell(\nu+\ell+1)_\tau}\nonumber \frac{(a+\delta)_{\ell+\tau}(b+\delta)_{\ell+\tau}(a+\delta-\nu)_\tau(b+\delta-\nu)_\tau}{\Gamma(k) \Gamma(k-\tau) ({d \over 2}+k-1)_\ell (k-{d \over 2}-\ell-\tau+1)_\tau}  \, .
\eea
Now we have
\bea
r^{(2)}(k, \ell, \tau+k) = - R^{a,b}_2(k, \ell) \times c^{a, b}_{\tau, \ell} \, ,
\eea
which leads to $R(k, \ell)$ for this case, 
\bea
R^{a,b}_2(k, \ell) &=&  - { r^{(2)}(k, \ell, \tau+k) \over c^{a, b}_{\tau, \ell} }\cr
&=& -\frac{ k (-1)^k}{(k!)^2} \frac{\left({d\over 2} -1-k\right)_{2k}
\prod_{i=\pm a,\pm b}\left( {\ell +{d\over 2} -k\over 2}+ i \right)_{k}}{\left(\ell+{d\over 2} -1-k\right)_{2k} \left(\ell+{d\over 2} - k\right)_{2k}}   \, .
\label{eq:residue2}
\eea
Finally, we will consider the third case, $\Delta_k=d+\ell +k -1$. 
Now the residue is 
\bea
r^{(3)}(k, \ell, \tau) = \frac{(-1)^{\ell + \tau + k}(2\nu)_\ell}
{\tau!\ell!(\nu)_\ell(\nu+\ell+1)_\tau}\nonumber \frac{\prod_{i=a,b}
(i+\delta)_{\ell+\tau}(i+\delta-\nu)_\tau}{ \Gamma(k) \Gamma(\tau-k) (d+\ell +k -2)_\ell({d \over 2}+\ell +k -\tau-1)_\tau }
 \, .
\eea
From 
\bea
r^{(3)}(k, \ell, \tau+k) = - R^{a,b}_3(k, \ell+k) \times c^{a, b}_{\tau-k, \ell+k} \, ,
\eea
we have finally,
\bea
 R^{a,b}_3(k, \ell) = -\frac{ k (-1)^k}{(k!)^2}\frac{(\ell+1-k)_{k}  \left( {1-k \over 2 } + a\right)_{k}  \left( {1-k \over 2 } + b\right)_{k}   }{(\ell+ {d \over 2}-k)_{k}} \, .
\label{eq:residue3}
\eea
It is well known that poles in a conformal block associated with a primary
$\OO_\ell$  occur at special scaling dimensions and spin $[\D_k',\ell']$ 
where some descendant of  the state created by $\OO_\ell$ becomes null 
\cite{Kos:2014bka, Penedones:2015aga}, which generalizes the original results of Zamolodchikov for two-dimensional conformal blocks~\cite{Zamolodchikov:1985ie}. This null state and its descendants form 
together a sub-representation, thus the residue of the associated pole is proportional to a conformal block, as we found in (\ref{eq:position&residue}).
It is interesting to notice that requiring the cancellation of singularities in the conformal block expansion of $u^\delta$ reproduces exactly the 
complete list of null states and their residues  of any CFT in arbitrary space dimensions. Actually our table 
\ref{tab:polepositions} coincides exactly with a similar table in
\cite{Kos:2014bka,Penedones:2015aga} and our residues (\ref{eq:residue1}),(\ref{eq:residue2}) and (\ref{eq:residue3}) coincide, apart  form a different normalization of the conformal blocks, with those calculated there with a completely different method. 

\subsection{Singularities of conformal blocks and higher-spin theories}
It should not be surprising that the analytic properties of generic conformal blocks are intimately connected with the rich and nontrivial structure of higher-spin gauge theories \cite{Vasiliev:2014vwa} that underlies  generic free CFTs and, as we will see, dictates also their nearby critical points. The first and third sets of dimensions shown in table \ref{tab:polepositions} are in one-to-one correspondence with the dimensions of the partially (or better: {\it multiply}) conserved higher-spin currents and their corresponding generalized Killing tensors of generic non-unitary theories, such as those of scalars with kinetic terms $\Box^{k}$ \cite{Brust:2016gjy,Brust:2016zns,Brust:2016xif,Bonifacio:2016blz}, but also fermionic ones that have not been studied yet. Indeed, in such theories one can construct partially conserved higher-spin currents $J_{\mu_1 \ldots  \mu_\ell}$ which are symmetric and traceless spin-$\ell$ operators satisfying equations such as
\be
\label{eq:pcurrents}
\partial_{\mu_1}\cdots\partial_{\mu_{\ell-t}}J_{\mu_1 \ldots \mu_\ell}=0\,,\,\,\,t=0,1,2, \ldots ,\ell-1\,.
\ee
 The dimensions of these currents are $\Delta_{\ell,t}=d-1+\ell -(\ell-t)$. The positive integer $t$ is the {\it depth} of partial conservation and the maximal depth case $t=\ell-1$ corresponds to the usual conserved higher-spin currents with dimensions $\Delta_{\ell,\ell-1}=d-2+\ell$. We then observe that $\D_{\ell,t}$  coincide with the positions of the singularities in the third line of table 1 if we identify $\ell-t=k$. When interactions are turned-on one generally expects that the conservation equation (\ref{eq:pcurrents})  is modified as
\be
\label{eq:pcurrentsnq}
\partial_{\mu_1}\cdots\partial_{\mu_{\ell-t}}J_{\mu_1 \ldots \mu_\ell}\sim {\cal O}_{\m_{\ell-t+1} \ldots \m_\ell}
\ee
The RHS represent operators with spin $\ell'= \ell-k$ and dimensions $\Delta'_{\ell',t}=d+\ell'+k-1$ which  coincide with the dimension and spin of  the would-be null states in the third line in table 1. 

The existence of the above partially conserved currents is  connected with existence of corresponding generalized Killing tensors. These have been discussed extensively in \cite{Boulanger:2012dx, Joung:2012hz, Joung:2012rv,Bekaert:2012vt,Basile:2014wua,Bekaert:2013zya,Grigoriev:2014kpa,Joung:2015jza} as well as more recently in \cite{Brust:2016gjy,Brust:2016zns,Brust:2016xif,Bonifacio:2016blz}. For our limited purposes here, however, it is simpler to just generalize the discussion of the usual higher-spin conserved currents in \cite{Giombi:2013yva}. Consider the marginal deformations of the form
\be
\label{eq:hssource}
\int J_{\m_1 \ldots \m_\ell}h_{\m_1 \ldots \m_\ell}\,.
\ee
Following \cite{Giombi:2013yva} one expects that under relatively mild assumptions, such as the existence of a large-$N$ expansion, $h_{\m_1 \ldots \m_\ell}$ can be considered as  spin-$\ell$ {\it partially massless} gauge field in a possible nontrivial UV fixed point (induced partially massless higher-spin gauge theory) of the theory. The existence of the gauge fields $h_{\m_1 \ldots \m_\ell}$ implies the presence of corresponding Killing tensors that generalize the conformal Killing equation for gravity. For example, in the case of the usual conserved higher-spin currents and their corresponding higher-spin gauge fields, the Killing equation takes the form \cite{Giombi:2013yva}
\be
\label{eq:killingeq}
(\hat{L}^{t=\ell-1}\cdot v)_{\m_2 \ldots \m_\ell}\equiv \partial_{(\m_1} v_{\m_2 \ldots \m_\ell)}-\frac{\ell-1}{d+2\ell-4}g_{(\m_1\m_2}\partial_\nu v_{\m_3 \ldots \m_\ell\n)}=0
\ee
where the parentheses denote total symmetrization and trace subtraction. Notice that in this case, to a spin-$\ell$ gauge fields corresponds a spin-$(\ell{-}1)$ Killing tensor. For partially conserved higher-spin currents of depth $t$ one expects a generalization of (\ref{eq:killingeq}) with more derivatives, such that to a spin-$\ell$ partially massless gauge field of depth $t$  corresponds a generalized Killing tensor of spin $\ell'=\ell-(\ell-t)\equiv \ell-k$. From (\ref{eq:hssource}) we can read the dimensions of $h_{\m_1 \ldots \m_\ell}$ to be $\tilde{\Delta}_{\ell,t}=1-2\ell+t\equiv 1-\ell+k$ and these coincide with the dimensions of the would-be null states in the first line of table 1. The Killing equation (\ref{eq:killingeq}) and its generalization sets to zero the unphysical gauge degrees of the free partially massless gauge field $h_{\m_1 \ldots \m_\ell}$. When the extended higher-spin gauge symmetry is broken, however, one generally expects that the theory is no longer free\footnote{The arguments for the case of the usual higher-spin gauge theories were presented e.g. in \cite{Maldacena:2011jn,Maldacena:2012sf}, but one expects that they generalize to the case of partially massless higher-spin gauge theories as well.} and those degrees of freedom enter the spectrum of the interacting theory. Schematically one can write 
\be
\label{eq:dh}
H_{\m_1 \ldots \m_\ell}=h_{\m_1\ldots\m_\ell}+(\hat{L}^{t}\cdot v)_{\m_{s-t+1}\ldots \m_s}
\ee
where the spin-$\ell$ field $H_{\m_1\ldots\m_\ell}$ is no longer a partially massless gauge field. This is of course a sketch of the expected Higgsing mechanism for partially massless gauge fields, or equivalently the corresponding multiplet recombination.  From the above we can read the dimension of the spin $\ell'=\ell-(\ell-t)\equiv \ell-k$ Killing tensors to be $\tilde{\Delta}'_{\ell',t}=1-\ell'-k$. We then recognise these operators as the singularities in the first line of table 1. 
Finally, the second set of dimensions in table \ref{tab:polepositions} corresponds to states and their {\it shadows} irrespective of their spin. The generalized multiplet recombination that we are describing here is the explicit realization of the algebraic analysis of \cite{Boulanger:2012dx, Joung:2012hz, Joung:2012rv,Bekaert:2012vt,Basile:2014wua,Bekaert:2013zya,Grigoriev:2014kpa,Joung:2015jza}. 

\section{Generalized free field theories and their central charges}
\label{sec:GFCFT}
Before studying generalized Wilson-Fisher fixed points, namely interacting theories, we will briefly review the generalized free field theories in the context of conformal bootstrap \cite{Guerrieri:2016whh}. Scalar generalized free conformal  field theories (GFCFTs) can be defined as the CFTs  generated 
by a single elementary scalar field $\phi$ with scaling dimension 
$\delta$ and 2pt function  normalized to be 
$\bra \phi(x_1)\phi(x_2)\ket=1/x_{12}^{2\delta}$.
All other correlation functions of the theory, either of $\phi$ or its composites as well as currents built from $\phi$, are given by
simple Wick contractions. As a consequence all the correlation functions with an odd 
number of $\phi$'s vanish, a condition that may be called {\it elementariness}. The simplest nontrivial example of a correlation function is the 4pt function of the elementary 
fields which is given by
\beq
\bra \phi(x_1)\phi(x_2)\phi(x_3)\phi(x_4)\ket=\frac{g(u,v)}{x_{12}^{2\delta}
x_{34}^{2\delta}},\;\;g(u,v)=1+u^\delta+\left(\frac uv\right)^\delta\,.
\label{eq:gf4pt}
\eeq
Taking advantage of the expansion of $u^\delta$ given in (\ref{eq:cbexpansion})
and putting $a=b=0$, it is easy to obtain the conformal block expansion 
of $g(u,v)$. It suffices to note that the exchange $x_1\leftrightarrow x_2$ in (\ref{eq:gf4pt}) 
entails $u\leftrightarrow u/v$ and a change of sign of the 
$G_{\D,\ell}(u,v)$ with $\ell $ odd, while those with $\ell$ even stay unchanged, then
\beq
g(u,v)=1+2\sum_{\tau=0}^\infty\sum_{n=0}^\infty
\lambda_{\delta,\tau,2n}^{0,0}G_{2\delta+2\tau+2n,\ell=2n}(u,v),
\label{eq:gfexpansion}
\eeq
where the $\lambda$'s are defined in (\ref{eq:cbexpansion}) and 
(\ref{eq:cab}). This result coincides with the one found in 
\cite{Fitzpatrick:2011dm} using Mellin space methods.

As $\delta$ is an arbitrary (real) number, generalized free CFTs do not necessarily admit a Lagrangian 
description. For the latter to be true one would require a non-vanishing coupling 
of the theories to the energy momentum tensor $T_{\mu\nu}$. According to (\ref{eq:gfexpansion}) 
the subclass coupled to $T_{\mu\nu}$, i.e. to the primary of scaling dimension 
$\D=2\delta+2\tau+\ell=d$ and spin $\ell=2$ has $\delta=d/2-k$ with 
$k=1,2,\dots$;
$k=1$ corresponds to the canonical free theory, while all cases with $k>1$ describe non-unitary theories as the fundamental scalar $\phi$ lies below the unitarity bound. According to table \ref{tab:polepositions} the dimension of this 
field coincides with the subclass of scalar  states having a null scalar 
descendant. A specific property of the conformal block expansion 
(\ref{eq:gfexpansion}) in this subclass of theories, as first noted in 
\cite{Guerrieri:2016whh}, is that some of the OPE coefficients are singular 
for some space dimension $d$ depending on the scale dimensions of $\phi$. 
Since the expanded function $g(u,v)$ is regular these singularities must cancel with corresponding singularities of the conformal blocks. This corresponds exactly to the cancellation mechanism that has been explained in full generality in the previous section.
   
A particular class of generalized free CFTs coupled to the  energy momentum tensor appears to play an important
role in the study of the $1/N$ expansions of vector-like theories. One quite 
intriguing observation made in \cite{Diab:2016spb} is that the highly 
nontrivial results for the $1/N$
 corrections to the {\it central charges}\footnote{By that we colloquially mean
the coefficient in front of the 2pt function of the energy momentum tensor.}  of vector-like theories in $d\geq 6$ simplify 
considerably when $d$ is even. There, the nontrivial result is
simply given by the sum of two terms, each one of them being the 
contribution of a {\it free} CFT. The general formula can be presented as
\be
\label{cTgen}
C_{T}^{(s,f)}(d)=NC_{T}^{(\phi,\psi)}(d)+c_T^{(\sigma_2,\sigma_1)}(d)+O(1/N)\,,\,\,\,d=2
n\,,\,n=2,3,4,5, \ldots
\ee
The case with $d=4$ is relevant only for the fermions. The first term is the 
contribution of the free vector-like CFTs of $N$ canonical scalars $\phi$ or 
$N$ Dirac spinors  $\psi$ in $d$ dimensions, with
\be
\label{CTb0}
C_T^{(\phi)}(d)=\frac{d}{(d-1)}\,, \,\,\,\,C_T^{(\psi)}(d)
=\frac{d}{2}\Tr\mathbb{I} \, ,
\ee
where the corresponding  spinor representation has dimension 
$\Tr\mathbb{I}$. The second term in (\ref{cTgen}) is the contribution of 
the free CFT of a  {\it single} generalized free field, the  $\sigma$ field. 
In the scalar case 
(for $d\geq 6$) this is the non-unitary $\sigma_2$ scalar with fixed 
scaling dimension $\Delta_2=2$ in any $d$, while in the fermionic case it is the $\sigma_1$ scalar with fixed dimension $\Delta_1=1$. 
In a Lagrangian description $\sigma_2$ and $\sigma_1$ can be described by higher derivative 
actions.
The explicit results read
\be
\label{ctcd}
c_T^{(\sigma_2)}(d)=\frac{(-1)^{\frac{d}{2}+1}d(d-4)(d-2)!}{(d-1)
\left(\frac{d}{2}+1\right)!\left(\frac{d}{2}-1\right)! }\,,\,\,\,\,\,
c^{(\sigma_1)}_T(d)=\
\frac{(-1)^{\frac{d}{2}}d(d-2)(d-2)!}{2\left(\frac{d}{2}+1\right)!
\left(\frac{d}{2}-1\right)! } \, .
\ee

The observation above indicates that for even dimensions $d\geq 4,6$ 
(for fermions and scalar respectively), there exist two types of free 
CFTs that are naturally connected to each other. On the one hand we have 
the canonical free CFTs of scalars $\phi$ and fermions $\psi$ whose 
corresponding scaling dimensions are
$\Delta_\phi=\frac{d}{2}-1$ and $\Delta_\psi=\frac{d}{2}-\frac{1}{2}$. 
On the other hand, we have the associated $\sigma$CFTs of $\sigma_2$ and 
$\sigma_1$.
The latter can be consistently defined as generalized free CFTs in any even
dimension. Their OPE structure was studied in \cite{Diab:2016spb} and their
corresponding central charges were calculated to be exactly (\ref{ctcd}),
either by using the OPE or by the direct evaluation of their energy momentum
tensor \cite{Osborn:2016bev}.

It is also interesting to note that having at hand the general results (\ref{eq:cbexpansion}) and (\ref{eq:cab}) we can give a general formula for the central charge of the whole class of scalar GFCFTs that are coupled to the energy momentum tensor. The result follows using the fact that the coefficient in front of the energy momentum conformal block is determined by a Ward identity \cite{Petkou:1994ad}, and then taking properly into account the normalization of the conformal blocks that we use here (see e.g. \cite{Paulos:2014vya}) we have that
\be
\label{eq:ct}
2\lambda^{0,0}_{\frac{d}{2}-k,k-1,2}=\left(\frac{d}{2}-k\right)^2\frac{C^{(\phi)}_T(d)}{c_T^{(k)}(d)}
\ee
From (\ref{eq:cbexpansion}) and (\ref{eq:cab}) we then obtain
\beq
c_T^{(k)}(d)={\rm cos}[(k-1)\pi]\frac{k\Gamma(\frac{d}{2}-k+1)\,\Gamma(\frac{d}{2}+k+1)}{\Gamma(\frac{d}2+2)\,\Gamma(\frac{d}2)} C_T^{(\phi)}(d)\, ,
\label{ct}
\eeq
where clearly we have $C_T^{(\phi)}(d)\equiv c_T^{(1)}(d)$. 

For a given even $d$ we may  define the normalized {\it total central charge } of the theory as the sum of the  central charges for all free scalars with $\Delta_k=\frac{d}{2}-k\,,k=1,2,3,..$.  Except for $k=1$, these are all nonunitary operators which in general correspond to ghost states. This is a delicate process because for even $d$ we hit the poles of $\Gamma(\frac{d}{2}-k+1)$ in the numerator of (\ref{ct}). However, if we {\it first } do the sum for general $d$, Mathematica gives a remarkably simple answer 
\beq
\label{eq:sumc}
C_T^{total}(d)\equiv \frac{1}{c_T^{(1)}(d) }\sum^{\infty}_{k=1} c_T^{(k)}(d)= {d \over 4(d+3) }\, ,
\eeq
Notice that this is positive despite the fact that the underlying theory contains negative norm states. Now we could take the $d=$even and obtain a finite result. Apparently, this result involves an underlying regularization that we do not yet understand. Nevertheless the result is consistent with the observation that 
\be
\label{eq:limc}
\lim_{d\rightarrow\infty}\frac{c_T^{(k)}(d)}{c_T^{(1)}(d)}=(-1)^{k-1}k\,.
\ee
In other words, in the limit $d\rightarrow \infty$  the sum (\ref{eq:sumc}) becomes the Euler alternating sum $1-2+3-4 \ldots$ which  is evaluated to $1/4$ after  appropriate regularization. This is still positive despite having summed over an infinity of ghost states. It would be interesting to unveil the possible physical interpretation behind this simple result.


\section{The smooth deformations of generalized free CFTs}
\label{sec:deformations}
After introducing the generalized free CFTs in the previous section, we now define and study their smooth deformations using conformal invariance as our only input. The results obtained include the calculation of the anomalous dimensions of an infinite 
class of scalar operators as well as OPE coefficients at the first 
non-trivial order in the $\epsilon$-expansion. Our approach is  similar in spirit to the one of Rychkov and Tan \cite{Rychkov:2015naa}, even if we do not use the equations of motion to obtain our results.
This way of reasoning suggests a possible definition 
of Wilson-Fisher fixed point and its extension to generalized free theories 
without any dynamical notion related to Lagrangians.

Let us start by considering  a generalized free 
CFT in $d$ dimensions.
We say that this  theory is close to a WF fixed point in $d-\epsilon$ dimensions if 
it admits a smooth deformation in $\epsilon$, i.e. if there is a one-to-one 
mapping to  another CFT in which any local operator $\OO_f$ of the free 
theory corresponds to 
an operator $\OO$ of the deformed theory with the same spin, but with scaling dimension and relevant 3pt couplings analytic functions of $\epsilon$ yielding the free results in the $\epsilon\rightarrow 0$ limit
\beq
\Delta_{\OO}(\epsilon)=\Delta_{\OO_f}+\gamma^{(1)}_{\OO}\epsilon+\gamma^{(2)}_{\OO}\epsilon^2+O(\epsilon^3)
; \quad \lambda_{\OO^i\OO^j\OO^k}(\epsilon)=
\lambda_{\OO_f^i\OO_f^j\OO_f^k}+O(\epsilon)\,,
\eeq
where $\gamma^{(i)}_{\OO}$ is the anomalous dimension of $\OO$ at the $i-$th order in the $\epsilon$ expansion.
Some, but not necessarily all,  of the above $\epsilon$-corrections should be different from zero. Note that the above definition does not imply that  all 
primary operators of the free theory correspond to  primary operators of the interacting one. Actually, the main ingredient of our calculation is the fact that some operator which is primary in free theory becomes a 
descendant when the interaction is turned on. 

\subsection{A simple example}\label{subsection:canonicald4}
We begin by applying in detail our method in the simple case of the deformation of a canonical free theory in $d=4-\epsilon$ dimensions before generalizing it to  the wide class of generalized free CFT with $\delta=\frac d2 -k$.
 Consider the following OPE in a free theory 
 \beq
[\phi_f]\times[\phi^2_f]=\sqrt{2}[\phi_f]+\sqrt{3}[\phi^3_f]+{\rm spinning ~operators} \, .
\label{eq:freet}
\eeq
The  OPE coefficients are computed  using  Wick contractions with normalization $[\phi^n] \equiv \phi^n/\sqrt{n!}$. Using the above, we calculate the mixed  4pt function $\bra\phi_f(x_1)\phi_f^2(x_2)\phi_f(x_3)\phi_f^2(x_4)\ket$ in the form (\ref{4pt}) and obtain\footnote{From now on we will suppress for simplicity the $u,v$ dependence of the conformal blocks.}
\beq
g_f(u,v)=2 \, G^{a_f,b_f}_{\Delta_{\phi_f},0}+ 3 \, G^{a_f,b_f}_{\Delta_{\phi_f^3},0} + {\rm spinning ~blocks} \, ,
\label{eq:freeexp}
\eeq 
where $\Delta_{\phi_f}=\frac d2-1$, $\Delta_{\phi_f^3}=3\Delta_{\phi_f}$, and 
\be
\label{eq:afbf}
a_f=-b_f=-\frac{\Delta_{\phi_f}-\Delta_{\phi^2_f}}{2}=-\frac{d-2}4
\ee
Notice that  according to (\ref{eq:position&residue}) and the table \ref{tab:polepositions} the conformal block $G^{a_f,b_f}_{d/2-1,0}$ appears to be singular having a simple pole exactly at $\D=\frac d2 -1$, however one can 
see that the corresponding residue $R^{a_f,b_f}(1,0)$ computed in (\ref{eq:residue2}) is zero. 
Let us investigate the viability of a deformed theory by setting
\beq
\D_{\phi^n}=\D_{\phi_f^n}+\gamma^{(1)}_{\phi^n}\epsilon+\gamma^{(2)}_{\phi^n}\epsilon^2+O(\epsilon^3)
\label{eq:deformation}
\eeq
where $\gamma^{(i)}_{\phi^n}\equiv\gamma^{(i)}_n$ denotes the anomalous dimension
of $\phi^n$ at $i$-th order in $\epsilon$. 
In the deformed (thus interacting) theory the first contributing conformal
 block is
$G^{a,b}_{\D_{\phi},0}$ with 
\be
\label{eq:ab}
a=-b=-\frac{d-2}4+
\frac{\gamma^{(1)}_{\phi^2}-\gamma^{(1)}_{\phi}}2\,.
\ee
Now the residue is no longer vanishing:
\beq
R^{a,b}(1,0)=\frac{(d-2)(\gamma^{(1)}_{\phi^2}-\gamma^{(1)}_{\phi})^2}{4d}\,\epsilon^2+O(\epsilon^3),
\label{eq:resd4}
\eeq
thus eq. (\ref{eq:position&residue}) becomes
\beq
G^{a,b}_{\Delta_\phi,0}=\frac{R^{a,b}(1,0)
}{\D_\phi-\D_{\phi_f}}\,G^{a,b}_{\D_{\phi_f}+2,0}+\dots
\label{eq:desc}
\eeq
It follows that in an interacting theory with $\gamma^{(1)}_{\phi^2}-
\gamma^{(1)}_\phi\not=0$ there exists a scalar with scaling 
dimension $\D_{\phi_f}+2$ which is a descendant of $\phi$. 
Precisely the primary $\phi$ contains a sub-representation (a null state) 
with the same Casimir eigenvalue ( apart from $\epsilon$-corrections) and  
scaling dimension $\D_{\phi_f}+2$. It is important to stress that according to 
(\ref{eq:deformation})  this mechanism holds true only in the interacting 
theory; in the free theory the primary $\phi_f$ does not contain that 
sub-representation, so  eq.(\ref{eq:desc}) may be viewed as playing the analog of the  classical equation of motion of the Wilson-Fisher fixed point. 

Let us analyze the deformed theory at $d=4-\epsilon$. There are two cases 
to be considered:
\begin{itemize}
\item[a)] If we take $\gamma^{(1)}_{\phi}\not=0$ we would have
\beq
G^{a,b}_{\D_\phi,0}=\epsilon\frac{(d-2)(\gamma^{(1)}_{\phi^2}-
\gamma^{(1)}_{\phi})^2}
{4d\gamma^{(1)}_{\phi} } G^{a,b}_{\D_{\phi_f}+2,0}+{\rm finite~terms}.
\label{eq:wrong}
\eeq
At $d=4-\epsilon$  the descendant operator $\OO_{\D_{\phi_f}+2}$ of the deformed 
theory becomes degenerate with $\phi^3$, therefore, according with the 
assumed one-to-one mapping between free and deformed theory, also  
$\phi^3$ is a descendent and its contribution should match in the 
$\epsilon\to0$ limit with the coefficient of  $G^{a_f,b_f}_{\frac{3d}2-3,0}$ 
in (\ref{eq:freeexp}), however this is impossible because of 
the different dependence in $\epsilon$. 

Hence the only consistent case is:   
\item[b)] $\gamma^{(1)}_{\phi}=0$, when
\beq
G^{a,b}_{\D_\phi,0}=\frac{(d-2)(\gamma^{(1)}_{\phi^2})^2}{4d\gamma^{(2)}_{\phi} } 
G^{a,b}_{\D_{\phi_f}+2,0}+F^{a,b}_{\D_\phi}.
\label{eq:subrap}
\eeq
and  $F^{a,b}_{\D_\phi}$ forms in the limit $\epsilon\to 0$   another 
eigenfunction of $C_2$ having the same eigenvalue with $G^{a,b}_{\D_{\phi_f}+2,0}$.\footnote{
It is worth noting that the decomposition (\ref{eq:subrap}) in two terms 
survives in the limit $\epsilon\to0$, while  the first term is absent if 
we directly put  $\epsilon=0$ in (\ref{eq:resd4}).} Now the coefficient of the descendant is finite, hence there is a 
perfect matching with the $\phi^3$ contribution in the free theory if and only if
\beq
\frac{\left(\gamma^{(1)}_{\phi^2}\right)^2}{\gamma^{(2)}_{\phi}}=12 \, ,
\label{eq:prediction4}
\eeq
which is consistent with the classic results for the corresponding anomalous 
dimensions evaluated using other methods (see e.g.  \cite{Zinn-Justin:2002}).
\end{itemize}

Let us stress that eq.s (\ref{eq:subrap}) and (\ref{eq:prediction4}) 
rigorously show, using no other assumptions than conformal invariance,  that the only consistent smooth deformation of the canonical free theory in $d=4-\epsilon$ 
is an interacting theory in which  
$\phi^3$ is a descendant of $\phi$. Clearly, this corresponds to the interacting
theory generated by the marginal $\phi^4$ perturbation and the conformal block expansion of the deformed theory becomes
\beq
g_I(u,v)=(2+O(\epsilon)) G^{a,b}_{\D_{\phi},0}  + {\rm spinning ~ blocks}.
\label{eq:deformed}
\eeq
 
Now, new results on OPE
coefficients can be 
obtained by considering the deformations of suitable OPEs in which  
a $\phi^3_f$ contribution appears on the RHS, for instance
\beq
[\phi_f^2]\times[\phi_f^5]=\sqrt{10}[\phi_f^3]+5\sqrt{2}[\phi_f^5]+\sqrt{21}
[\phi_f^7]+{\rm spinning~operators} \, ,
\eeq   
or
\beq
[\phi_f]\times[\phi_f^4]=2[\phi_f^3]+\sqrt{5}[\phi_f^5]+{\rm spinning~operators} \, .
\eeq
Both the above OPE expansions  contain a $\phi^3_f$ contribution which is no longer a primary field in the deformed theory, hence it should be 
replaced by the conformal block of $\phi$. Repeating the  
calculation described previously and using the corresponding values for the residues  we obtain 
the following OPE coefficients
\begin{align}
\lambda_{\phi^2\phi^5\phi}^2&=5\gamma^{(2)}_\phi\epsilon^2+O(\epsilon^3)=
\left(\frac{\sqrt{5}}{2\sqrt{3}}\gamma^{(1)}_{\phi^2}\epsilon\right)^2+O(\epsilon^3);\\
\lambda_{\phi\phi^4\phi}^2&=2\gamma^{(2)}_\phi\epsilon^2+O(\epsilon^3)=
\left(\frac1{\sqrt{6}}\gamma^{(1)}_{\phi^2}\epsilon\right)^2+O(\epsilon^3)\,,
\end{align}
where the second equalities are obtained using (\ref{eq:prediction4}).

\subsection{The $\phi^{2n}$ critical points}\label{subsection:phi2n}
In this section we will apply the method explained in detail above to the much more general class of generalized free CFTs in various spacetime dimensions. Furthermore, we will demonstrate that we are able calculate precise values of anomalous dimensions instead of their ratios as in (\ref{eq:prediction4}). We begin by considering the free OPE, but for a more general class of operators, 
\be
\label{pp1}
[\phi^p_f] \times [\phi^{p+1}_f]=\sum_{n=1}^{p+1} \lambda_{p,p,2n-1}
[\phi^{2n-1}_f]+\ldots \, .
\ee
The corresponding OPE coefficients are calculated to be 
\be
\lambda_{p,p,2n-1}=B_{2n-1,n}\sqrt{\frac{p+1}{(2n-1)!}}(p-n+2)_{n-1}\,,
\ee
with $B_{n,m}$ the binomial coefficient. We can now take 
in general $\Delta_{\phi_f}=\frac{d}{2}-k,\,k=1,2,3, \ldots$, which correspond the subclass of 
generalized free CFTs coupled to the energy-momentum tensor, as discussed in section 
\ref{sec:GFCFT}. Inserting (\ref{pp1}) into the direct channel of the 4pt 
function $\bra\phi^p\phi^{p+1}\phi^p\phi^{p+1}\ket$  one obtains an 
expansion in terms, among others, of scalar conformal blocks of the type 
$G^{a_f,-a_f}_{\Delta_{\phi^{2n-1}_f,0}}$ with $a_f=\Delta_{\phi_f}/2$. When we 
smoothly deform the theory, each OPE coefficient should be modified with a term which vanishes in the limit of $\epsilon \rightarrow 0$. Most importantly 
the operator $\phi^{2n-1}$ becomes a descendant, which should be removed from 
the expansion. As we discussed in the previous subsection, the conformal 
block in the interacting theory $G^{a,-a}_{\Delta_\phi,0}$ (now for the interacting theory $a=a_f + \gamma_{{p+1}}-\gamma_{p}$ and $\Delta_\phi=\Delta_{\phi_f} + \gamma_{1}$) has a pole 
with a residue proportional to the conformal block $G^{a,-a}_{{d \over 2}+k,0}$ 
which is precisely the missing conformal block of the operator $\phi^{2n-1}$ 
in the free theory in the limit $\epsilon \rightarrow 0$. 
Matching the operator dimensions requires the spacetime dimension to be $d=2nk/(n-1)$. One finds again 
for $k=1$ and $n=2$ the special case treated in the previous subsection. 

To be explicit, the four-point correlator $\langle \phi^p(x_1)\, \phi^{p+1}(x_2) \, \phi^p(x_3)\, \phi^{p+1}(x_4) \rangle $ admits the following the conformal block expansion 
\bea
g_f(u,v) &=& \lambda^2_{p,p,1} G^{a_f, -a_f}_{\Delta_{\phi_f}, 0} + \lambda^2_{p,p,2n-1} G^{a_f, -a_f}_{\Delta_{\phi^{2n-1}_f}, 0} + \ldots  \, ,\\
g_I(u,v) &=& (\lambda^2_{p,p,1} + \mathcal{O}(\epsilon ) ) G^{a, -a}_{\Delta_{\phi}, 0} + \ldots \, ,
\eea
where $g_f(u,v)$ and $g_I(u,v)$ represent correlators in the free and interacting theories, respectively. Using then the explicit result for the residue in (\ref{eq:residue2}), the matching condition gives
\be
(p+1)\frac{(\gamma_{{p+1}}^{(1)}-\gamma_{p}^{(1)})^2}{\gamma_{1}^{(2)}}\frac{(-1)^{k+1}\left( \frac{d}{2} -k\right)_k}{4k\left( \frac{d}{2} \right)_k}=\lambda^2_{p,p,2n-1} \, ,
\ee
where we have also used the fact that $\lambda^2_{p,p,1} = p+1$. We see that the consistency condition requires $\gamma^{(1)}_{1}=0$ and this 
leads to  the following recursion relation
\be
\gamma_{{p+1}}^{(1)}-\gamma_{p}^{(1)}= \kappa(k,n) (p-n+2)_{n-1}\, ,
\ee
with 
\be
\kappa^2(k,n)=\frac{(-1)^{k+1}4 k \,B^2_{2n-1,n}\left(\frac{n k}{n-1} \right)_k}{(2n-1)!\left(\frac{k}{n-1}\right)_k}\gamma_1^{(2)} \,,
\ee
where we used $d=2nk/(n-1)$. There is seemingly a sign ambiguity here, however we will fix $\kappa(k,n)$ completely shortly. Since we have of course $\gamma_{0}=0$, the solution of the recursion relation 
can be written in the form
\be \label{eq:solution}
\gamma_{p}^{(1)}= \frac{\kappa(k,n)}{n}(p-n+1)_n.
\ee
Now the crucial observation is that the operator $\phi^{2n-1}$ that becomes a 
descendant of $\phi$ has a fixed dimension, namely ${d \over 2} + k$ as shown 
in table \ref{tab:polepositions}, therefore in the perturbed theory in 
$d-\epsilon$ we have
\beq
\D_{\phi^{2n-1}}=(2n-1)(\frac {d-\epsilon}2-k)+\gamma^{(1)}_{2n-1}\epsilon+O(\epsilon^2)=
{d-\epsilon \over 2} + k
\eeq
 From this fact we find explicitly 
its anomalous dimension to be $\gamma^{(1)}_{2n-1}=n-1$.
Plugging this into (\ref{eq:solution}) fixes $\kappa(k,n) = {n(n-1) \over (n)_n}$, which is interestingly $k$-independent. As promised, there is no sign ambiguity, and it yields
\bea \label{eq:finding1}
\gamma_{p}^{(1)}= \frac{(n-1)}{(n)_n}(p-n+1)_n \, .
\eea
From the result of $\kappa(k,n)$, we also obtain the anomalous dimension of $\phi$ at order $\epsilon^2$ as  
\bea \label{eq:finding2}
\gamma^{(2)}_1 = (-1)^{k+1} 2 { n \left( {k \over n-1} \right)_k \over k \left( {n \, k \over n-1} \right)_k } (n-1)^2 \left[ { (n!)^2 \over (2n)!} \right]^3 \, .
\eea
A few comments are in order. In the case of $k=1$, namely for the canonical scalar, $\gamma^{(2)}_1\equiv\gamma^{(2)}_\phi$ reduces to  
\beq
 \gamma^{(2)}_\phi=  2(n{-}1)^2 \left[ { (n!)^2 \over (2n)!} \right]^3\,,
\label{eq:gamma2}
\eeq 
which is a well-known multicritical result \cite{Zinn-Justin:2002}. 
In particular for $n=2$ eq.s (\ref{eq:finding1}) and (\ref{eq:gamma2}) 
correspond to the  $\phi^4$ theory in $d=4-\epsilon$ as we discussed briefly in previous section, while for $n=3$ these results describe  the interacting $\phi^6$ theory in $d=3-\epsilon$. More 
generally when $k>1$, we are able to describe smooth deformations of scalar generalized CFTs 
with $\Delta_{\phi}={d \over 2}-k={k \over n-1}$ in $d={2nk \over n-1}$. 
For $k>1$, we have assumed that we only turn on one possible marginal 
deformation of the form $\phi^{2n}$, in principle one may have marginal 
interactions with derivatives. Notice also that $k>1$ allows us to study 
multicritical, but non-unitary, theories in integer dimensions $d>6$.

\subsection{OPE coefficients}
\label{OPE_coefficients}
We can apply the same procedure described above to compute non-trivial OPE coefficients at the generalized Wilson-Fisher critical points. We begin by considering the general OPE in the free theory, 
\bea
[\phi^p] \times [\phi^{p+2m+1}] = \sum_{n=m+1}^{p+m+1} \lambda^f_{p, p+2m+1,2n-1} [\phi^{2n-1}]+\dots \, ,
\eea
with the free OPE  coefficients $\lambda^f_{p, p+2m+1,2n-1} $ given by
\bea
\lambda^f_{p, p+2m+1,2n-1}  = { \sqrt{\Gamma(2n) \Gamma(p+1) \Gamma(2m+p+2) } \over  \Gamma(n-m) \Gamma(n+m+1) \Gamma(m-n+p+2)  } \, .
\eea
When $m=0$, it reduces to eq.(\ref{pp1}). For the theory in  $d={2nk \over n-1}$ dimensions  with marginal deformation $\phi^{2n}$, we are now able to compute the following OPE coefficients at the interacting WF fixed points. 
Consider the 4pt correlator $\langle \phi^p \, \phi^{p+2m+1} \,  \phi^p \, \phi^{p+2m+1} \rangle$ with $m>0$ in free and interacting theory in terms of conformal block expansions, we have
\bea
g_f(u,v)&=& (\lambda^f_{p,p+2m+1, 2n-1})^2 \, G^{a_f,b_f}_{\Delta_{\phi^{2n-1}}}  + \ldots \, , \\
g_I(u,v)&=& (\lambda_{1,p,p+2m+1})^2 \, G^{a,b}_{\Delta_{\phi}}  + \ldots \, .
\eea
Note when $m>0$, the operator $\phi$ does not appear $g_f(u,v)$, but it should in general arise in the interacting theory. Thus the residue of the singular block $G^{a,b}_{\Delta_{\phi}}$ in $g_I(u,v)$ should reproduce $G^{a_f,b_f}_{\Delta_{\phi^{2n-1}}}$ in the free theory, which leads to the matching condition that reads
\bea
C_{1,p,p+2m+1} \times {R^{a,b}(k,0) \over \gamma^{(2)}_{\phi} \epsilon^2 } = { \Gamma(2n) \Gamma(p+1) \Gamma(2m+p+2)  \over  \Gamma^2(n-m) \Gamma^2(n+m+1) \Gamma^2(m-n+p+2) } \, ,
\eea
here we denote $C_{1,p,p+2m+1} = (\lambda_{1,p,p+2m+1})^2$. The residue $R^{a,b}(k,0)$ is given by
\bea\label{Rab}
R^{a,b}(k,0) = { (-1)^{k+1} \left[((m+1)(d/2-k))_{k} \right]^2 \left[(m(k-d/2))_{k} \right]^2 \over k! (k-1)! (d/2-k)_{2k} } \, .
\eea
Thus we obtain an infinite class of non-trivial OPE coefficients of the theory, 
\bea
C_{1,p,p+2m+1}
&=& 
\epsilon^2    \left[ { (n!)^3 \over (2n)!} \right]^2
{ (n-1)^2 \Gamma(p+1) \Gamma(2m+p+2)  \over  \Gamma^2(n-m) \Gamma^2(n+m+1) \Gamma^2(m-n+p+2) } \cr
&\times& 
 { \left( {k \over n-1} \right)_k \over \left( {n \, k \over n-1} \right)_k } {  \Gamma^2(k)  \, ({k \over n-1})_{2k}  \over  \left[({(m+1)k \over n-1})_{k} \, ({-m k \over n-1})_{k} \right]^2 } \, ,
\eea
for any $p$ and $m \geq 1$. Here we have plugged in the result of the anomalous dimension (\ref{eq:finding2}) from the previous section. In the case of canonical scalar theory with $k=1$ the OPE coefficient simplifies to
\bea
C_{1,p,p+2m+1}
&=& 
\epsilon^2   \left[ { (n!)^3 \over (2n)!} \right]^2
{ (n-1)^4 \Gamma(p+1) \Gamma(2m+p+2)  \over \left[ \Gamma(n-m) \Gamma(n+m+1) \Gamma(m-n+p+2) m (m+1)\right]^2 } \, .
\eea
To our knowledge, this general result of OPE coefficients is new. For some special cases, for instance those computed in~\cite{Nii:2016lpa}, we find agreement (after taking into account the difference of normalizations). 

%
%

\section{$O(N)$ invariant  models}\label{subsection:O(N)}

In this Section, we will apply our method to theories with global symmetries in general dimensions. We focus on theories with global O$(N)$ symmetries having elementary scalars fields $\phi_i$, $i=1,2, \ldots, N$ that transform as vectors under $O(N)$, with dimensions $\Delta_{\phi_i} = d/2-k$. In contrast to the single scalar case, composite operators involving many $\phi_i$'s fall into rank-$s$ symmetric tensor representations of $O(N)$. We will show here that using our method we will be able to calculate the anomalous dimensions  as well as various non-trivial OPE coefficients for all those operators in tensors representations of $O(N)$ at the WF fixed points of the theories. Explicit results for $d=4k$, $d=3k$ and $d=8k/3$ are given and several other examples can be found in Appendix \ref{freeOPEs} and \ref{determination}.

\subsection{Anomalous dimensions}
The correlators that will provide relevant information about anomalous dimensions are the following two types of mixed 4pt functions: 
\begin{align}
\langle [\phi_i\phi^{2(p-1)}](x_1) \, [\phi^{2p}](x_2) \,
[\phi_i\phi^{2(p-1)}](x_3)  \, 
[\phi^{2p}](x_4)\rangle &= \frac{1}{(x^2_{12}x^2_{34})^{(2p-\frac{1}{2})\Delta_\phi}}\left(\frac{x^2_{13}}{x^2_{24}}\right)^{\frac{\Delta_\phi}{2}}g^{(1)}(u,v) \, ,\\
\langle [\phi^{2p}](x_1)  \, [\phi_i\phi^{2p}](x_2)  \,
[\phi^{2p}](x_3)  \, 
[\phi_i\phi^{2p}](x_4)\rangle&= \frac{1}{(x^2_{12}x^2_{34})^{(2p+\frac{1}{2})\Delta_\phi}}\left(\frac{x^2_{13}}{x^2_{24}}\right)^{\frac{\Delta_\phi}{2}}g^{(2)}(u,v) \, .
\end{align}
Here $\phi^{2p}\equiv(\phi_i\phi_i)^p$ denotes the $O(N)$ singlet. The conformal block expansions of the above 4pt functions in the free and interacting theory are 
\begin{align}\label{gfgi}
g_f^{(m)}(u,v)&= (\lambda^{(m)}_{p,p,0})^2 G^{a_f,-a_f}_{\Delta_{\phi_f},0}+\dots+ (\lambda^{(m)}_{p,p,s})^2 G^{a_f,-a_f}_{\Delta_{(2n+1)\phi_f},0}+\dots \\
g_I^{(m)}(u,v)&=\left( 
(\lambda^{(m)}_{p,p,0})^2+\mathcal{O}(\epsilon) \right)G^{a,-a}_{\Delta_{\phi},0}+\dots,
\end{align}
where $m=1,2$, while $\lambda^{(1)}_{p,p,n}=\lambda_{p,p-1,n}$ and  $\lambda^{(2)}_{p,p,n}=\lambda_{p,p,n}$.
The definition of the free OPE coefficients $\lambda_{p_1,p_2,p_3}$ and the recursion relations used to compute them for general $p$ can be found 
in the Appendix~\ref{freeOPEs}.

As before, deforming the free theory spectrum and OPE coefficients, the operator $[ \phi_i \phi^{2n} ]$ 
becomes a descendant of $\phi_i$ for $d=2k(n+1)/n-\epsilon$. The matching conditions for both 4pt  functions above become
\begin{align}
r(k,n)\frac{(\gamma_{p-1,1}^{(1)}-\gamma_{p,0}^{(1)})^2}{\gamma_{0,1}^{(2)}}=\left(\frac{\lambda_{p,p-1,s}}{\lambda_{p,p-1,0}}\right)^2 \, ,
\quad r(k,n)\frac{(\gamma_{p,0}^{(1)}-\gamma_{p,1}^{(1)})^2}{\gamma_{0,1}^{(2)}}=\left(\frac{\lambda_{p,p,n}}{\lambda_{p,p,0}}\right)^2 \, ,
\label{eq:tworelations}
\end{align}
where we have defined
\be
r(k,n)=\frac{(-1)^{(k+1)}\left(\frac{k}{n}\right)_k}{4k\left(k\frac{n+1}{n}\right)_k} \, .
\ee
Here we denote the anomalous dimension of $\phi^{2p}$ as $\gamma_{p,0} =\gamma^{(1)}_{p,0} \epsilon + \gamma^{(2)}_{p,0} \epsilon^2 + \ldots $, and correspondingly  for $[\phi_i \phi^{2p}]$ as
$\gamma_{p,1} =\gamma^{(1)}_{p,1} \epsilon + \gamma^{(2)}_{p,1} \epsilon^2 + \ldots $. 
Removing $\gamma^{(1)}_{p,0}$ from (\ref{eq:tworelations}), we obtain the following  recursion relation for $\gamma^{(1)}_{p,1}$
\be
\gamma^{(1)}_{p,1}-\gamma^{(1)}_{p-1,1}=\left(\frac{\lambda_{p,p,n}}{\lambda_{p,p,0}}+\frac{\lambda_{p,p-1,n}}{\lambda_{p,p-1,0}}\right)\sqrt{\frac{\gamma_{0,1}^{(2)}}{r(k,n)}} :=f(p,n)\sqrt{\frac{\gamma_{0,1}^{(2)}}{2^{2n}n!(N/2+1)_n\, r(k,n)}}\, .
\label{recursion}
\ee
We have fixed the sign ambiguity in taking the square root by matching the formulae for $N=1$ to the corresponding ones in the single scalar case. The function $f(p,n)$ is determined in terms of the OPE coefficient $\lambda_{p,q,n}$, whose explicit formula can be found in Appendix~\ref{gamma01} up to $n=6$. Solving then  the recursion relation (\ref{recursion}), we obtain the general formula for the anomalous dimension, 
\be\label{ONrec}
\gamma_{p,1}^{(1)}=g(p,n)\sqrt{\frac{\gamma_{0,1}^{(2)}}{2^{2n}n!(N/2+1)_n \,r(k,n)}} \, ,
\ee
where $g(p,n)$ up to $n=6$ are given in Appendix~\ref{gamma01}. As before $[\phi^i {\phi^{2n}}]$ becomes a descendant with a fixed conformal dimension $d/2+k=k(2n+1)/n$ and  this leads to $\gamma^{(1)}_{n,1}=n$. This result allows us to fix $\gamma_{0,1}^{(2)}$ in (\ref{ONrec}), 
\bea \label{eq:finding3}
\gamma_{0,1}^{(2)}=\frac{n^2}{g^2(n,n)}2^{2n}n!\left(\frac{N}{2}+1\right)_n\,r(k,n)\, .
\eea
Plugging this result back to~\eqref{ONrec} and~\eqref{eq:tworelations} we also obtain, 
\bea \label{eq:finding4}
\gamma^{(1)}_{p,1} = \frac{n}{g(n,n)}g(p,n) \, , \quad\quad\quad\quad \gamma^{(1)}_{p,0}=\gamma_{p,1}-\frac{\lambda_{p,p,n}}{\lambda_{p,p,0}}\frac{n}{g(n,n)\sqrt{2^{2n}n!(N/2+1)_n}}.
\eea
Explicit formulas for $\lambda_{p,q,n}$ and $g(p,n)$ can also be found in Appendix~\ref{gamma01} and Appendix~\ref{freeOPEs}, respectively. 

The above analysis can be extended to the general class of symmetric traceless rank-$s$ $O(N)$ tensors operators defined as such as $\phi_{i_1} \phi_{i_2}\dots\phi_{i_s} \phi^{2p} -\text{traces} \equiv\phi^{(s)}_{i_1\dots i_s}\phi^{2p}$. As concrete examples we give here the rank-$2$ and the rank-$3$ tensors
\bea
\phi^{(2)}_{ij}\phi^{2p} &=&\phi_i \phi_j \phi^{2p} -{1\over N} \delta_{ij} \phi^{2(p+1)} \, , \\ 
\phi^{(3)}_{ijk}\phi^{2p} &=& \phi_i \phi_j\phi_k \phi^{2p} -\frac{1}{N+2} (\delta_{ij}\phi_k+\delta_{ik}\phi_j+\delta_{jk}\phi_i) \phi^{2(p+1)} \, . 
\eea
To study the anomalous dimension of rank-$s$ tensor operator as $\gamma^{(1)}_{p,s}$ we consider the correlator 
\be
\langle [\phi^{(s)}_{i_1 \ldots i_{s-1}k}] \,\phi^{2(p-1)}](x_1)\, [\phi^{(s-1)}_{i_1 \ldots i_{s-1}}\phi^{2p}](x_2)\, [\phi^{(s)}_{j_1 \ldots j_{s-1}k}\phi^{2(p-1)}](x_3)\, [\phi^{(s-1)}_{j_1 \ldots j_{s-1}}\phi^{2p}](x_4)\rangle.
\ee
The above correlator can be expanded exactly as in~\eqref{gfgi} and the matching 
condition yields the following recursion relation, 
\be \label{eq:recursion-s}
(\lambda^{(s,s-1,1)}_{p-1,p,0})^2 r(k,s)\frac{(\gamma^{(1)}_{p-1,s}-\gamma^{(1)}_{p,s-1})^2}{\gamma_{0,1}^{(2)}}=(\lambda^{(s,s-1,1)}_{p-1,p,n})^2\,. 
\ee
There is again a sign ambiguity which is fixed by considering some suitable formal $N$-limits. We first note that in the formal limit $N \rightarrow 0$ the rank-$2$ tensor operators above effectively becomes a scalar since the trace part dominates. Hence its anomalous dimension should reduce to that of a scalar for $N\rightarrow 0$. Similarly, the rank-$3$ tensor reduces to a vector operator in the formal limit $N \rightarrow -2$. More generally the formal limit $N \rightarrow -2(s-2)$ reduces a rank-$s$ tensor operator to a rank-$(s-2)$ tensor. In this way, not only we fix the sign ambiguity, but can also perform highly non-trivial checks of our results. 

Taking then the square root in (\ref{eq:recursion-s}) we find
\be\label{eq:spin-j}
\gamma^{(1)}_{p-1,s}=\gamma^{(1)}_{p,s-1}-\sqrt{\frac{\gamma_{0,1}^{(2)}}{r(k,n)}}\frac{\lambda^{(s,s-1,1)}_{p-1,p,0}}{\lambda^{(s,s-1,1)}_{p-1,p,n}} \, ,
\ee
where the free OPE coefficients $\lambda^{(s,s-1,1)}_{p_1,p_2,p_3}$ and the corresponding recursion relations used to compute them in general are given in Appendix~\ref{gamma01}. We can now give a few explicit examples of the general results above. We consider  the cases $n=1,2,3$, correspondingly $d=4k, 3k, 8/3k$: 
\begin{itemize}
\item The $n=1$ case corresponds to considering the deformed theories in $d=4k-\epsilon$.
The anomalous dimension $\gamma_{0,1}^{(2)}$ of the $\phi_i$ is  given by~\eqref{eq:finding3}
\be
\gamma_{0,1}^{(2)}=\frac{8(N+2)}{g^2(1,1)}r(k,1)=\frac{(-1)^{k+1}(k)_k}{2k(2k)_k}\frac{(N+2)}{(N+8)^2},
\ee
and substituting in~\eqref{eq:finding4} yields
\be
\gamma^{(1)}_{p,1}=\frac{p(N+6p+2)}{N+8},\quad\quad\gamma^{(0)}_{p,0}=\frac{p(N+6p-4)}{N+8}.
\ee
Using $\gamma^{(1)}_{p,1}$ we can solve the recursion relation in~\eqref{eq:spin-j}
\be
\gamma^{(1)}_{p,s}=\frac{s(s-1)+p(N+6(p+s)-4)}{N+8},
\ee
which yields $\gamma^{(1)}_{p,0}$ and $\gamma^{(1)}_{p,1}$ for $s=0$ and $s=1$.

\item The $n=2$ case corresponds to considering theories in $d=3k-\epsilon$.
Again plugging $n=2$ in~\eqref{eq:finding3} we get
\be
\gamma_{0,1}^{(2)}=\frac{32(N+2)(N+4)}{g^2(2,2)}r(k,2)=\frac{(-1)^{k+1}(k/2)_k}{8k(3k/2)_k}\frac{(N+2)(N+4)}{(3N+22)^2},
\ee
\be
\gamma^{(1)}_{p,1} = { 10 p + 3N+2 \over 3 (3N+22)} p (2 p-1),\quad\quad \gamma^{(1)}_{p,0} = { 2(10 p + 3N-8) \over 3 (3N+22)} p (p-1).
\ee
Applying the recursion relation in~\eqref{eq:spin-j}, we obtain the general formula for the anomalous dimension for tensor operators, 
\be
\gamma_{p,s}^{(1)}=\frac{(2p+s-2)(s(s-1)+p(3N+10(p+s)-8))}{3(3N+22)} \, .
\ee

\item Finally we consider $n=3$ in $d=8k/3-\epsilon$
\be
\gamma_{0,1}^{(2)}=\frac{(-1)^{k+1}3(k/3)_k}{k(4k/3)_k}\frac{(N+2)(N+4)(N+6)}{(1072+3N(50+N))^2} \,
\ee
and 
\begin{align}
\gamma_{p,s}^{(1)}&=\frac{1}{3N(N+50)+1072}\left[s^4 
+ 2 (10 p - 3)s^3
+  (90 p^2 + 6 (N - 19) p + 11) s^2  \right. \cr
&+  2 (70 p^3 + 15 (N - 11) p^2 - (18 N - 97) p - 3) s
\cr
&
\left. + \, p (p - 1) \left(10 (3 N - 17) p + 70 p^2 + 3 ({ N^2 \over 2} - 15 N + 32) \right)  \right] \, .
\end{align}

\end{itemize}
Our results for $\gamma^{(2)}_{0,1}$ when $k=1$ and for $n=1,2,3$ agree with older known results obtained with completely different methods~\cite{Hofmann:1991ge,Hager:2002uq}.  Finally, our results for $\gamma_{p,s}^{(1)}$ in all three cases satisfy  the following highly non-trivial relationship 
\bea
\gamma_{p,s}^{(1)} \, = \, \gamma_{p+1,s-2}^{(1)} \, ,
\eea
which is obtained taking the formal limits $N \rightarrow -2(s-2)$ as discussed above. 

\subsection{OPE coefficients}
As in the case of a single scalar in section~\ref{OPE_coefficients}, an infinite set of non-trivial OPE coefficients can be computed by considering a class 
of 4-point functions
\bea
\langle [\phi^{2p}](x_1) \, [\phi_i\phi^{2(p+m)}](x_2) \, [\phi^{2p}](x_3) \, [\phi_i\phi^{2(p+m)}](x_4)\rangle \, .
\eea 
For $m=1,\dots,s$, the free OPE $[\phi^{2p}]_f\times [\phi_i\phi^{2(p+m)}]_f$ contains the operator $[\phi_i\phi^{2n}]_f$. 
Since when we smoothly deform the theory $[\phi_i\phi^{2n}]$ becomes a descendant of $\phi_i$,  it follows that it cannot appear as a primary in the interacting OPE. This gives 
 the matching condition
\be\label{ONOPE}
\lambda^2_{p,p+m,0}\frac{R^{a,b}(k,0)}{\gamma_{0,1}^{(2)}}=\lambda^2_{p,p+m,n},
\ee
where $\lambda^2_{p,p+m,0}$ is zero in the free theory and proportional to $\epsilon^2$ in the deformed theory. Here we list the explicit results for the  OPE coefficients for the cases $n=1,2,3$. 
\begin{itemize}
\item The $n=1$ case corresponds to consider the deformed theories in $d=4k-\epsilon$. The non-trivial OPE coefficients $\lambda_{p,p+1,0}$ can be obtained from~\eqref{ONOPE}
\be
\lambda^2_{p,p+1,0}=\frac{[\Gamma(2k)]^2(k)_k}{2k\,k!\,\Gamma(3k)(2k)_k}\frac{(p+1)(N+2p)(N+2p+2)}{N(N+8)^2} \epsilon^2 \, .
\ee

\item The $n=2$ case corresponds considering theories in $d=3k-\epsilon$. In this case there are two different sets of OPE coefficients that can be computed looking at the OPE $[\phi^{2p}]\times [\phi_i\phi^{2p+2}]$
and $[\phi^{2p}]\times [\phi_i\phi^{2p+4}]$
\begin{align}
\lambda^2_{p,p+1,0}&=\frac{k!\,\Gamma(k)(k/2)_k(k/2)_{2k}}{8k\,[(-k/2)_k\,(k)_k]^2(3k/2)_k}\frac{25 p^2(p+1)(N+2p)(N+2p+2)}{N(3N+22)^2} \epsilon^2 \, ,\\
\lambda^2_{p,p+2,0}&=\frac{k!(k-1)!(k/2)_k(k/2)_2k}{16k\,[(-k)_k]^2[(3k/2)_k]^3}\frac{(p+1)(p+2)(N+2p)(N+2p+2)(N+2p+4)}{N(3N+22)^2} \epsilon^2 \, .
\end{align}

\item Finally we consider the $n=3$ case for $d=8k/3-\epsilon$. The OPE coefficients for this case can again be easily computed using results given in the Appendix~\ref{determination}, yielding
\begin{align}
\lambda^2_{p,p+1,0}&=\frac{p^2(p+1)(N+2p)(N+2p+2)(N+14p-8)^2[\Gamma(-k/3)\Gamma(4k/3)\Gamma(k+1)]^2}{2N(1072+3N(N+50))^2[\Gamma(k/3+1)\Gamma(5k/3)]^2} \epsilon^2,\\
\lambda^2_{p,p+2,0}&=\frac{49\pi \,4^{1-2k}p^2(p+1)(p+2)(N+2p)(N+2p+2) [\Gamma(-2k/3)\Gamma(4k/3)\Gamma(k)]^2}{N(1072+3N(N+50))^2[\Gamma(k/3)]^4[\Gamma(k+1/2)]^2} \epsilon^2,\\
\lambda^2_{p,p+3,0}&=\frac{(p{+}1)(p{+}2)(p{+}3)(N{+}2p)(N{+}2p{+}2)(N{+}2p{+}4)(N{+}2p{+}6)k!(k-1)!(k/3)_k(k/3)_{2k}}{2k\,N(1072+3N(N+50))^2[(-k)_k]^2[(4k/3)_k]^3} \epsilon^2 \, .
\end{align}

\end{itemize}

%
%

\section{Multiple deformations} \label{subsection:d=6}
So far we have only considered theories with only one marginal deformation. In this section, we consider smooth deformations of free CFT's with more than one marginal deformations. We show that we can still use our method, but  some new and interesting features arise. As an example 
we consider a theory in $d=6- \epsilon$ with two scalars denoted as $\sigma$ and $\phi_i$ with the latter being the $O(N)$ vector of the previous section. In this particular case, the possible marginal deformations are $\sigma^3$ and  $\sigma \phi_k^2$. Due to $O(N)$ symmetry, as the interaction turns on we can have $\sigma \phi_i$ becoming a descendant of $\phi_i$, while a linear combination, $\mathcal{O}_d = g_1 \sigma^2 + g_2 \phi^2 \, ,$
becomes a descendant of $\sigma$. This implies that there is another linear combination of $\sigma^2$ and $\phi^2$ which remains as a primary, namely
\be
\label{eq:Op}
\mathcal{O}_p ={1 \over \sqrt{2N} \sqrt{g_1^2 N + g_2^2} } (g_1 N \sigma^2 - g_2 \phi^2 )\,.
\ee
Here $\langle \mathcal{O}_p\, \mathcal{O}_d \rangle = 0$ and $\langle \mathcal{O}_p\, \mathcal{O}_p \rangle  = 1$. We then expand the 4pt function $\langle \phi_i \phi_i \phi_j \phi_j \rangle$, in terms of conformal blocks to obtain 
\bea
g_f(u,v) &=& 1+ {2 \over N} \, G_{\Delta_{\phi^2},0} + \ldots \\
g_I (u,v)&=& 1+ C^2_{\phi_i \phi_i \sigma} \, G_{\Delta_{\sigma},0}  + {2 \, g^2_2 \over g_1^2 N^2+g_2^2 N } G_{\Delta_{\mathcal{O}_p},0} +\ldots \,,  \nonumber
\eea
where again $g_f(u,v)$ denotes the free theory expansion while $g_I(u,v)$ the interacting one. 
In the free theory $\phi^2$ is a primary operator, but after the interaction is turned on only $\OO_p$  remains as a conformal primary. Then, taking the  $\epsilon \rightarrow 0$ limit,  we must require that $G_{\Delta_{\mathcal{O}_p}, 0}$ becomes $G_{\Delta_{\phi^2}, 0}$. On the other hand, due to the singularity structure of conformal blocks, we have,  
\bea
G_{\Delta_{\sigma}, 0} = {1 \over 6 ( \Delta_{\sigma} - \Delta_{\sigma_f}   ) } G_{\Delta_{\phi^2}, 0}+ \ldots \, ,
\eea
namely $G_{\Delta_{\sigma}, 0}$ produces another contribution to $G_{\Delta_{\phi^2}, 0}$ in the free theory limit. 
Thus, matching with the free theory yields
\bea
{2 \, g^2_2 \over g_1^2 N^2+g_2^2 N } + {C^2_{\phi_i \phi_i \sigma} \over 6  \gamma^{(1)}_{\sigma} \epsilon} = {2 \over N} ,\, 
\eea
and we obtain
$C^2_{\phi_i \phi_i \sigma} = {12 g_1^2 \over g_1^2N + g_2^2} \gamma^{(1)}_{\sigma} \epsilon \, .$ 
Moreover, from the mixed 4pt function  $\langle \phi_i \sigma \phi_i \sigma \rangle$, again in the conformal block expansion we have, 
\bea
g_f(u, v) &=&1+ \, G_{\Delta_{\phi_i \sigma}, 0} + \ldots \cr
g_I(u, v) &=&1+ C^2_{\phi_i \phi_i \sigma} \, G_{\Delta_{\phi_i}, 0}  + \mathcal{O}(\epsilon) G_{\Delta_{O_p}, 0} + \ldots \, ,
\eea
where the same OPE coefficient $C^2_{\phi_i \phi_i \sigma}$ appears in the interacting theory. From the singularity of $G_{\Delta_{\phi_i}, 0}$, we see that $\phi_i \sigma$ is descendant of $\phi_i$, and the matching condition leads to $C^2_{\phi_i \phi_i \sigma} = 6 \gamma^{(1)}_{\phi_i}$. One may then cancel out the OPE coefficient $C^2_{\phi_i \phi_i \sigma}$ to obtain the ratio of the anomalous 
dimensions as
\bea \label{eq:finding8}
{ \gamma^{(1)}_{\phi} \over \gamma^{(1)}_{ \sigma }} ={2 g_1^2 \over g_1^2 N + g_2^2 } \, .
\eea
We can obtain more information by considering other correlators. For instance from the 4pt function 
$\langle \sigma \sigma  \sigma  \sigma \rangle$, a similar analysis leads to, 
\bea \label{eq:finding9}
C^2_{\sigma\sigma\sigma} =  {12 \, g_2^2 \over g_1^2N + g_2^2} \gamma^{(1)}_{\sigma} \epsilon \, .
\eea
Our results are consistent with the loop calculations of 
  \cite{Fei:2014yja} and also with the more recent work \cite{Nii:2016lpa}
 that uses the equations of motion.
 
It is also interesting to note that the correlation function  $\langle \phi_i  \, \sigma \,  \phi_i \, \sigma \rangle$ we considered earlier could be expanded in the crossing $\{1,3\}\{2,4\}$ channel. In this different choice of the OPE expansion, we then have 
\bea
g_f(u, v) &=& 1 \, , \cr
g_I(u, v) &=&1+C_{\phi_i \phi_i \sigma}  C_{\sigma\sigma\sigma} \, G_{\Delta_{\sigma}, 0}  - {2 g_1 g_2 \over g_1^2 N + g_2^2 } G_{\Delta_{O_p}, 0} + \ldots \, .
\eea
Note in this channel the free theory correlator $g_f(u, v)$ does not contain the operator $\phi^2$ which would the descendant of $\sigma$, thus it requires following cancellation, 
\bea
{ C_{\phi_i \phi_i \sigma}  C_{\sigma\sigma\sigma}   \over 6 \gamma^{(1)}_{\sigma} \epsilon} - {2 g_1 g_2 \over g_1^2 N + g_2^2 } = 0 \, .
\eea
This identity is indeed verified by using explicit results of $C_{\phi_i \phi_i \sigma}$ and $C_{\sigma\sigma\sigma}$, and thus shows the consistency of our analysis. 

\section{Conclusion} \label{sec:conclusion}
In this work we  have used properties of CFTs,  and in particular the analytic structure of generic  conformal blocks, to study the possible smooth 
deformations of generalized free CFTs  in arbitrary dimensions. In this way we could define and generalize the notion of Wilson-Fisher fixed points in terms of conformal invariant concepts, with no reference to the renormalization goup or to any Lagrangian approach.  The examples analyzed include general classes of multicritical points, $O(N)$ invariant theories as well as theories with multiple deformations. Combining the OPE structure with
 universal properties of certain scalar null states we 
derived, at the 
first non-trivial order in the $\epsilon$-expansion, the anomalous dimensions
of  infinite classes of scalar local operators as well as non-trivial OPE coefficients in the interacting theories. For theories with $O(N)$ global 
symmetry we were able to consider general symmetric traceless $O(N)$ tensor operators. In the particular cases where 
other computational methods were applied, the results agree. Our method allows us to put huge classes of critical theories under a unified calculation scheme
without using any kind of dynamical equation of motion. We also  remark that unlike the usual conformal bootstrap program, neither crossing symmetry nor unitarity have been used in our scheme. Therefore the method can be useful to study non-unitary theories that are relevant in the description of certain critical systems.  Nevertheless, it should be mentioned that going beyond leading order using our method would require additional input. Our method is equivalent to calculating leading order critical quantities using the tree-level results. The next-to-leading order calculations would require the equivalent of conformal block analysis for a one-loop graph, which would in turn require the contributions from an infinite class of conformal blocks. Alternatively, one might try to use the information coming from crossing symmetry. Clearly, this discussion is extremely interesting but beyond the scope of the present work.

There is a plethora of directions that one can follow based on this work and we briefly mention a few of them here. One could for example extend our methods to fermionic non-unitary theories in general dimensions and perhaps even try to identify possible supersymmetric WF fixed point of generalized free CFTs. An extension of our methods should probably work for matrix and tensor critical models, e.g. such as the one discussed recently in \cite{Klebanov:2016xxf}. The anomalous dimensions of partially conserved higher-spin currents is also another well-defined project within the range of applicability of our methods. The intimate connection of the conformal OPE with higher-spin gauge theory, whose surface we have just scratched here, is by itself an interesting subject and we believe that our methods provide a concrete and computationally solid path for its study. To this end, we also mention that our  result regarding the total central charge of non-unitary free CFTs in even dimension is reminiscent of the remarkable recent calculations for the partition function of higher-spin theories (see e.g. \cite{Beccaria:2015vaa,Beccaria:2016tqy}). Work on some of the paths mentioned above is already in progress.

\section*{Acknowledgements}
A large part of this work was performed while A. C. P. was visiting CHPT at Ecole Polytechnique which he wishes to thank for the excellent hospitality extended to him. A. C. P. wishes also to thank  X. Bekaert, K. Hinterbichler, P. M. Petropoulos, M. Picco, Z. Skvorstov, S. Sleight, M. Taronna and A. Tseytlin for useful discussions and correspondence. 
F. G. wishes to thank E. Brezin and S. Hikami for helpful discussions. This material is based upon work supported by the U.S. Department of Energy, Office of Science, Office of High Energy Physics, under Award Number DE-SC0010255. The work of A. C. P. is partially supported by the MPNS–COST Action MP1210 “The String Theory Universe”. The work of A.~L.~G. is funded under CUniverse research promotion project by Chulalongkorn University (grant
reference CUAASC).

\appendix
\section{Computing the conformal block expansion using Casimir operator} 
\label{appendix:expansion}
In this appendix we compute the conformal block expansion (\ref{eq:cbexpansion})
using the properties of the quadratic Casimir operator $C_2$. 
The conformal blocks are eigenfunctions of $C_2$:
\beq
C_2\, G_{\D,\ell}^{a,b}(u,v)=c_2(\D,\ell) G_{\D,\ell}^{a,b}(u,v),~~
c_2(\D,\ell)=\frac{\D(\D-2\nu+2)+\ell(\ell+\nu)}2.
\eeq
 The Casimir operator can be written, using the notation of \cite{Hogervorst:2013kva}, as
\beq
C_2={\cal D}_z+{\cal D}_{\bar{z}}+2\nu\frac{z\bar{z}}{z-\bar{z}}\left[(1-z)
\frac{\rm d}{{\rm d}z}-(1-\bar{z})\frac{\rm d}{{\rm d}\bar{z}}\right]
\eeq
with
\beq
 {\cal D}_z=(1-z)z^2\frac{{\rm d}^2}{{\rm d}z^2}-(a+b+1)z^2
\frac{\rm d}{{\rm d}z}-ab\,z
\eeq
and
\beq
u=z\bar{z};~v=(1-z)(1-\bar{z});~\nu=\frac d2-1;
 a=-\frac12\D_{12}^-;~b=\frac12\D_{34}^-.
\eeq
Our normalization is chosen in such a way that for $z=\bar{z}\to 0$ 
$G_{\D,\ell}^{a,b}(u,v)=z^\D+{\rm higher~order~terms}$. First, we want to show that the spectrum of conformal blocks contributing to  $u^\delta$ is
$\Sigma=\{\D=2\delta+2\tau+\ell,\ell\},~(\tau,\ell=0,1,\dots)$.
For this purpose we define the following differential operator
\beq
\Omega(n)=\prod_{[\D,\ell]\in\Sigma,2\tau+\ell< n}\left(C_2-c_2(\D,\ell)\right).
\label{eq:Omega}
\eeq
This operator projects out all the conformal blocks of $\Sigma$ belonging to the subset $2\tau+\ell<n$. It turns out, in the $z=\bar{z}\to0$ limit,
\beq
\Omega(n)\,u^\delta=O(z^{2\delta+n})
\eeq 
showing that no other conformal block  with $\D<2\delta+n$ can contribute.

In order to compute the coefficient $\lambda^{a,b}_{\delta,\tau_o,\ell_o}$ 
of the conformal block corresponding to the representation 
$[\D_o=2\delta+2\tau_o+\ell_o,\ell_o]\in \Sigma_n\equiv\Sigma\&\, (2\tau+\ell<n)$ it suffices to omit the factor corresponding to it in the product (\ref{eq:Omega}):
\beq
\Omega'_{\D_o,\ell_o}=\prod_{[\D,\ell]\in\Sigma_n,[\D,\ell]\not=[\D_o,\ell_o]}
\left(C_2-c_2(\D,\ell)\right).
\eeq
We have, again in the  $z=\bar{z}\to0$ limit,
\beq
\Omega'_{\D_o,\ell_o}\,u^\delta=z^{\D_o}\lambda^{a,b}_{\delta,\tau_o,\ell_o}
\prod_{[\D,\ell]\in\Sigma_n,[\D,\ell]\not=[\D_o,\ell_o]}
\left(c_2(\D_o,\ell_o)-c_2(\D,\ell)\right)+{\rm higher~order~terms}
\label{eq:lambda}
\eeq 
from which  $\lambda^{a,b}_{\delta,\tau_o,\ell_o}$ can be computed. It is important to note that the limit $z=\bar{z}\to0$ has to be taken only at the very end 
of the calculation. In practice 
we apply this procedure for low values of $n$, guess a formula, and 
check it to higher values of $n$. The result is reported 
in (\ref{eq:cbexpansion}) and (\ref{eq:cab}).

\section{Singular conformal blocks and the OPE}
To illustrate the role played by the singular conformal blocks in the conformal OPE, we consider the following 4pt function of a scalar $\phi$ with dimension $\Delta_\phi$
\be
\label{ex:4ptphi}
\langle\phi(x_1)\phi(x_2)\phi(x_3)\phi(x_4)\rangle = \frac{1}{(x_{12}^2x_{34}^2)^{\Delta_\phi}}g(u,v)
\ee
In free theory limit, using Wick contractions one then finds
\be
\label{ex:g}
g_f(u,v)  = 1+u^{\Delta_\phi}+\left(\frac{u}{v}\right)^{\Delta_\phi}
\ee
and the conformal block expansion is of the form
\be
\label{ex:gOPE}
g_f(u,v)  = 1+2 \,G_{\Delta_{\phi^2},0}+ \cdots
\ee
where the dimension of the operator $\phi^2$ is $2\Delta_\phi$. However, as we have not yet specified neither $\Delta_\phi$ not $d$, it is possible that we might encounter problems with the OPE (\ref{ex:gOPE}). Namely, suppose that the dimension of $\phi^2$ is such that the corresponding block is singular, i.e.  $2\Delta_\phi = \frac{d}{2}-1$\footnote{The argument generalizes straightforwardly also when $2\Delta_\phi=d/2-k$ with $k>1$.}. If we require that the free (\ref{ex:gOPE}) is regular, the singularity of the conformal block of $\phi^2$ must somehow cancel. We can see how this happens in the explicit case when $\Delta_\phi=2$ where for general $d$ the free OPE (\ref{ex:gOPE}) is found to be
\begin{align}
&g_f(u,v)=1+2\,G_{4,0}+\frac{24(d-1)}{5d}G_{6,2}+\frac{80(d^2-1)}{21(d+2)(d+4)}G_{8,4}+\dots\nonumber\\
&+\frac{8(d-6)}{d(d-10)}G_{6,0}+\frac{288(d-6)(d-1)}{7(d-14)d(d+4)}G_{8,2}+\frac{640(d-6)(d^2-1)}{11(d-18)(d+2)(d+4)(d+8)}G_{10,4}+\dots\nonumber\\
&+\frac{36(d-8)(d-6)^2}{(d-7)(d-12)(d-14)d(d+2)}G_{8,0}+\frac{320(d-8)(d-6)^2(d-1)}{(d-18)(d-16)d(d+4)(d+6)(d-7)}G_{10,2}+\dots \, .
\label{twotwo}
\end{align}
We note that some of the OPE coefficients become singular at certain spacetime dimension. For instance at $d=10$, the coefficient of conformal block $G_{6,0}$ becomes singular. It turns out that this singularity is precisely cancelled by the singularity of the conformal $G_{4,0}$ whose residue is proportional to $G_{6,0}$. Indeed, the state that corresponds to the conformal block $G_{4,0}$ is $\phi^2$, whose conformal dimension hits the singularity point as $\Delta=4= d/2-k$ with $k=1$ when $d=10$. On the other hand, the primary state contributes to $G_{6,0}$ is given by
\bea
S_{6,0}={i } \left( {1 \over 16} \partial_{\mu} \phi \partial^{\mu} \phi -
{\Delta_{\phi} \over 8 (2+2 \Delta_{\phi} -d)} \phi \partial^2 \phi \right) \, ,
\eea
as first studied in \cite{Guerrieri:2016whh}. For the special values of $\Delta_{\phi}=2$ and $d=10$, the coefficient simplifies, it becomes
\bea
S_{6,0} = {i \over 16} \left( \partial_{\mu} \phi \partial^{\mu} \phi + \phi \partial^2 \phi \right)
= {i \over 32} \partial^2 \phi^2 \,. 
\eea
Thus we see that the primary $S_{6,0}$ now becomes a descendant of $\phi^2$, as a primary and a descendant at the same time, it is thus a null state. Or more explicitly, since for this particular case $\Delta_{\phi^2}=d/2-1$, which is actually the dimension of a canonical scalar, thus its two-point function is killed by $\partial^2$.  Similar analysis applies to other perfect cancellations for other terms in eq. (\ref{twotwo}):  between the terms with $G_{4,0}$ and  $G_{8,0}$ in $d=12$; between the terms with $G_{6,2}$ and $G_{8,2}$   in $d=14$; between the terms with  $G_{8,4}$ and $G_{10,4}$ in $d=18$ etc.

On the other hand when we deform the theory we assume an OPE of the form
\be
\label{ex:g2OPEint}
g_I(u,v)  = 1+g_*G_{\Delta_{\phi},0}(u,v) + \cdots
\ee
As we have seen in the text, this implies a conformal cubic interaction which is relevant for the description of the Wilson-Fisher fixed  point of $\phi^3$ in $d=6-\epsilon$. Now if we want a smooth free field theory limit, we must be concerned for the singularity of the conformal block $G_{\Delta_\phi,0}(u,v)$ as $\Delta_\phi\rightarrow\frac{d}{2}-1$, and in fact the operator $\phi$ does not appear in the free OPE (\ref{ex:gOPE}). What happens now is that the critical coupling $g_*$ has a zero in the free theory limit which kills the regular leading term of $G_{d/2-1,0}(u,v)$ and cancels its pole that multiplies the descendant with dimension $\frac{d}{2}+1$. The result is that the descendant emerges as the primary operator $\phi^2$ in the free theory limit, and from the relations $d/2+1=2\Delta_\phi=d-2$ we find the critical dimension $d=6$ as expected. This is the mechanism described in great detail in the main text.

\section{Wick contractions and free OPE coefficients for $O(N)$ vector models}
\label{freeOPEs}

In order to compute some relevant OPE coefficients, we consider two classes of three-point functions 
involving scalars and vector operators
\begin{align}
G_1(p,q,s)&=\langle \phi^{2p}(x_1) \phi_i\phi^{2q}(x_2) \phi_i\phi^{2s}(x_3) \rangle\\
G_2(p,q,s)&=\langle \phi^{2p}(x_1) \phi^{2q}(x_2) \phi^{2s}(x_3) \rangle
\end{align}
Performing the first step of Wick contraction we can derive the following recursion relations
\begin{align}
G_1(p,q,s)&=\frac{4p q}{x_{12}^{4\Delta_\phi}} G_1(p{-}1,q{-}1,s)+\frac{2p (2s+N)}{x_{12}^{2\Delta_\phi}x_{13}^{2\Delta_\phi}}G_2(p{-}1,q,s)+\frac{(2s+N)}{x_{23}^{2\Delta_\phi}}G_2(p,q{-}1,s{-}1)\\
G_2(p,q,s)&=\frac{2p(2(q{-}1){+}N)}{x_{12}^{4\Delta_\phi}}G_2(p{-}1,q{-}1,s)+\frac{4ps}{x_{12}^{2\Delta_\phi}x_{13}^{2\Delta_\phi}} G_1(p{-}1,q{-}1,s{-}1)+\frac{2s}{x_{23}^{2\Delta_\phi}} G_1(p,q{-}1,s{-}1).
\end{align}
To recursively compute the free three-point functions we specify the initial conditions
\be
G_1(0,0,0)=\langle \phi_i(x_2)\phi_i(x_3)\rangle=\frac{N}{x_{23}^{2\Delta_\phi}},\quad\quad G_2(0,0,0)=\langle \mathbb{I} \rangle=1,
\ee
and
\begin{align}
G_1(p,0,p)&=\langle \phi^{2p}(x_1)\phi_i(x_2)\phi_i\phi^{2p}(x_3)\rangle=2^{2p}p! \left(\frac{N}{2}+1\right)_p \,\frac{N}{x_{23}^{2\Delta_\phi}x_{13}^{4p\Delta_\phi}}\\
G_2(p,0,p)&=\langle\phi^{2p}(x_1)\phi^{2p}(x_3)\rangle=2^{2p}p! \left(\frac{N}{2}\right)_p\,\frac{1}{x_{13}^{4p\Delta_\phi}}.
\end{align}
The OPE coefficients used in section~\ref{subsection:O(N)} are then defined as
\be
\lambda_{p,q,s}=\frac{\widetilde G_1(p,q,s)}{\sqrt{B_2(p)B_1(q)B_1(s)}} \, ,
\ee
where $\widetilde G_i$ is the structure constant of three-point functions and the normalizations of two-point functions $B_1(p)$ and $B_2(p)$ are defined as
\begin{align}
\langle \phi_i\phi^{2p}(x)\phi_i\phi^{2p}(0)\rangle&=\frac{2^{2p}p!\left(\frac{N}{2}+1\right)_p N}{(x^{2})^{(2p+2)\Delta_\phi}}=\frac{B_1(p)}{(x^{2})^{(2p+2)\Delta_\phi}}\\
\langle \phi^{2p}(x)\phi^{2p}(0) \rangle&=\frac{2^{2p}p!\left(\frac{N}{2}\right)_p }{(x^{2})^{2p\Delta_\phi}}=\frac{B_2(p)}{(x^{2})^{2p\Delta_\phi}} \, .
\end{align}
The OPE coefficients needed for the computation of the anomalous dimensions $\gamma_{0,1}^{(2)}$ are of the form $\lambda_{p,p,q}$
and $\lambda_{p,p-1,q}$, we show their values up to $q=3$: 
\begin{align}
&q=0: \quad\quad  \quad \lambda_{p,p-1,0}=\sqrt{\frac{2p}{N}}\quad\quad \lambda_{p,p,0}=\sqrt{\frac{2p+N}{N}}\, , \\
& q=1: \quad\quad  \quad \lambda_{p,p-1,1}=\sqrt{\frac{2p}{N}}\frac{N+6p-4}{\sqrt{2(N+2)}},\quad\quad \lambda_{p,p,1}=\sqrt{\frac{2p+N}{N}}\frac{3\sqrt{2}\,p}{\sqrt{N+2}} \, , \\
& q=2: \quad\quad  \quad 
\lambda_{p,p-1,2}=\sqrt{\frac{2p}{N}}\frac{2(p-1)(3N+10p-8)}{\sqrt{2(N+2)(N+4)}},\quad\quad \lambda_{p,p,2}=\sqrt{\frac{2p+N}{N}}\frac{\sqrt{2}\,p(N+10p-6)}{\sqrt{(N+2)(N+4)}} \, , \\
& q=3: \quad\quad  \quad  \lambda_{p,p-1,3}=\sqrt{\frac{p}{N}}\frac{(p-1)(3N(N+20p-30)+4(5p(7p-17)+48))}{\sqrt{6(N+2)(N+4)(N+6)}} \, , \\ \nonumber
& \quad\quad  \quad \quad\quad \quad \quad \lambda_{p,p,3}=\sqrt{\frac{2p+N}{N}}\frac{5p(p-1)(3N+14p-10)}{\sqrt{3(N+2)(N+4)(N+6)}} \, .
\end{align}
Here we now list relevant free OPE coefficients needed for the computation of $\lambda^2_{p,p+1,0}$, $\lambda^2_{p,p+2,0}$ and $\lambda^2_{p,p+3,0}$ in interacting theories (see eq.(\ref{ONOPE})), 
\begin{align}
&\lambda^2_{p,p+1,1}=\frac{(p+1)(N+2p)(N+2p+2)}{N(N+2)}\\
&\lambda^2_{p,p+1,2}=\frac{25p^2(p+1)(N+2p)(N+2p+2)}{N(N+2)(N+4)}\\
&\lambda^2_{p,p+2,2}=\frac{(N+2p)(N+2p+2)(N+2p+4){p+2 \choose 2}}{N(N+2)(N+4)}\\
&\lambda^2_{p,p+1,3}=\frac{3p{p+1\choose 2}(N+2p)(N+2p+2)(N+14p-8)^2}{N(N+2)(N+4)(N+6)}\\
&\lambda^2_{p,p+2,3}=\frac{49p(N+2p)(N+2p+2)(N+2p+4){p+2\choose 3}}{N(N+2)(N+4)(N+6)}\\
&\lambda^2_{p,p+3,3}=\frac{(N+2p)(N+2p+2)(N+2p+4)(N+2p+6){p+3\choose 3}}{N(N+2)(N+4)(N+6)}
\end{align}



\section{Functions $f(p,n)$, $g(p,n)$ and anomalous dimensions $\gamma_{0,1}^{(2)}$}
\label{determination}
\label{gamma01}
\begin{itemize}
\item $n=1$ case
\begin{align}
f(p,1)=N+12 p - 4 \, , \quad \quad\quad \quad g(p,1)=p(N+6p+2) \, .
\end{align}
The corresponding anomalous dimension given by
\bea
\gamma_{0,1}^{(2)}=\frac{(-1)^{k+1}(k)_k}{2k(2k)_k}\frac{N+2}{(N+8)^2} \, .
\eea

\item $n=2$ case   
\begin{align}
f(p,2)=4(N(4p-3)+4(2+p(5p-6)))\, , \quad\quad  g(p,2)=\frac{4}{3}p(2p-1)(3N+10p+2) \, .
\end{align}
The corresponding anomalous dimension given by
\bea
\gamma_{0,1}^{(2)}=\frac{(-1)^{k+1}(k/2)_k}{8k(3k/2)_k}\frac{(N+2)(N+4)}{(3N+22)^2}\, .
\eea

\item $n=3$ case
\begin{align}
&f(p,3)=2(p-1)(3N^2+90(p-1)N+8(5p(7p-11)+24)) \, , \\
&g(p,3)=p(p-1)(3N^2+30(2p-1)N+4(5p(7p-3)+2)) \, .
\end{align}
The corresponding anomalous dimension given by
\bea
\gamma_{0,1}^{(2)}=\frac{3(-1)^{k+1}(k/3)_k}{k(4k/3)_k}\frac{(N+2)(N+4)(N+6)}{(3N(N+50)+1072)^2} \, .
\eea

\item $n=4$ case
\begin{align}
&f(p,4)=16(p-1)(3N^2(3p-5)+14(2p-3)(4p-5)N+4(p(248+7p(9p-31))-96)) \, ,\\
&g(p,4)=\frac{8}{5}p(p-1)(2p-3)(15 N^2+70 (2p-1)N+4(7p(9p-5)-2)) \, .
\end{align}
The corresponding anomalous dimension given by
\bea
\gamma_{0,1}^{(2)}=\frac{(-1)^{k+1}(N+2)(N+4)(N+6)(N+8)(k/4)_k}{6k(3464+5N(3N+98))^2(5k/4)_k} \, .
\eea

\item $n=5$ case
\begin{align}
&f(p,5)=8(p-1)(p-2)(5N^3+10(28p-51)N^2+20(105p(p-3)+236)N\nonumber\\
&\quad\quad\quad\quad + 16(7p(3p(11p-42)+161)-480)) \, , \\
&g(p,5)=\frac{8}{3}p(p-1)(p-2)(5N^3+30N^2(7p-10)+20N(47+63p(p-2))\nonumber\\
&\quad\quad\quad\quad+24(2+7p(11+p(11p-24)))) \, .
\end{align}
The corresponding anomalous dimension given by
\bea
\gamma_{0,1}^{(2)}=\frac{(-1)^{k+1}15(N+2)(N+4)(N+6)(N+8)(N+10)(k/5)_k}{16k(139488+5N(3968+N(150+N)))^2(6k/5)_k} \, .
\eea

\item $n=6$ case
\begin{align}
&f(p,6)=32(p-1)(p-2)(5N^3(8p-21)+30N^2(147+2p(15p-67))  \\
&+4N(p(12761+297p(4p-23))-7770)+48(960+p(p(2338+11p(13p-87))-2482)))\, ,\nonumber \\
&g(p,6)=\frac{32}{7}p(p-1)(p-2)(2p-5)(35N^3+210N^2(3p-4)+28N(73+99p(p-2))\nonumber\\
&\quad\quad\quad\quad+24(2+p(169+11p(13p-30)))) \, .
\end{align}
The corresponding anomalous dimension given by
\bea
\gamma_{0,1}^{(2)}=\frac{(-1)^{k+1}9(N+2)(N+4)(N+6)(N+8)(N+10)(N+12)(k/6)_k}{320k\,(480576+7N(9796+5N(84+N)))^2(7k/6)_k} \, .
\eea

\end{itemize}

\bibliographystyle{abe}
\bibliography{bibliography}{}

\providecommand{\href}[2]{#2}\begingroup\raggedright
\end{document}